\documentclass[10pt,journal,letterpaper,twoside,twocolumn,nofonttune]{IEEEtran}

\usepackage{tabularx}

\usepackage{amsmath,amsfonts}
\usepackage{algorithmic}
\usepackage{algorithm}
\usepackage{array}
\usepackage{float}
\usepackage{textcomp}
\usepackage{stfloats}
\usepackage{url}
\usepackage{verbatim}
\usepackage{graphicx}
\usepackage{cite}
\usepackage{amsmath, xparse}
\usepackage[utf8]{inputenc}
\usepackage{textgreek}
\usepackage{xcolor}
\usepackage{cite}
\usepackage{amsmath,amssymb,amsfonts}
\usepackage{algorithmic}
\usepackage{graphicx}
\usepackage{textcomp}
\usepackage{bm}
\usepackage{float}
\usepackage{xcolor}
\usepackage{soul}
\usepackage{xfrac}
\usepackage{subfigure}

\usepackage{soul}

\definecolor{LightGray}{gray}{0.8}
\definecolor{Orange}{rgb}{1.0, 0.31, 0.0}
\definecolor{Green}{rgb}{0.3, 1.0, 0.3}
\definecolor{Blue}{rgb}{0.75,0.75,1}
\definecolor{Magenta}{rgb}{255, 0, 255}
\definecolor{Cyan}{rgb} {0, 255, 255}

\newcommand{\bea}{\begin{eqnarray}}
\newcommand{\beal}[1]{\begin{eqnarray}\label{#1}}
\newcommand{\eea}{\end{eqnarray}}
\def\balg#1#2\ealg{\begin{align}\label{#1}#2\end{align}}
\def\balgnl#1\ealgnl{\begin{align*}#1\end{align*}}

\ifCLASSINFOpdf
\else
\fi
\begin{document}
%
\title{Frequency-Domain Analysis of Wave Scattering by Spatially Dispersive Metasurfaces Using the Method of Auxiliary Sources}

%

\author{Minas~Kouroublakis, Nikolaos~L.~Tsitsas, \IEEEmembership{Senior Member, IEEE}, and Yehuda Leviatan, \IEEEmembership{Life Fellow, IEEE} 
\thanks{\emph{Corresponding author: Nikolaos~L.~Tsitsas}}
\thanks{Minas~Kouroublakis and Nikolaos~L.~Tsitsas are with the School of Informatics, Aristotle University of Thessaloniki, 54124 Thessaloniki, Greece
(e-mails: mkour2000@yahoo.com; ntsitsas@csd.auth.gr).
}
\thanks{Yehuda Leviatan is with Department of Electrical and Computer Engineering, Technion-Israel Institute of Technology, Haifa, 32000, Israel (e-mail: leviatan@technion.ac.il)}}

%
%

\markboth{IEEE TRANSACTIONS ON ANTENNAS AND  PROPAGATION}%
{IEEE TRANSACTIONS ON ANTENNAS AND  PROPAGATION}


%


\maketitle

\begin{abstract}
Spatially dispersive metasurfaces exhibit angle-dependent responses that cannot be accurately modeled using conventional local susceptibilities. Extended Generalized Sheet Transition Conditions (GSTCs) have been introduced to account for spatial dispersion by incorporating spatial derivatives of the electromagnetic fields. In this work, these extended GSTCs are integrated into the Method of Auxiliary Sources (MAS), resulting in a meshless simulation framework for the analysis of spatially dispersive metasurfaces. The proposed formulation is developed for infinite planar, finite planar, polygon shaped, and cylindrical metasurfaces, while it is general and also applicable to bianisotropic spatially dispersive metasurfaces. For validation, the numerical examples consider Lorentz-type spatial resonators, consistent with previously published studies. 
The extended GSTCs are enforced within the MAS via appropriate placement of auxiliary sources. Numerical results are presented for several geometries, including planar, polygonal, semicircular, and cylindrical metasurfaces. The obtained results show very good agreement with previously published data, demonstrating the accuracy and flexibility of the proposed method. The developed approach provides a simple and computationally efficient meshless alternative to existing techniques and constitutes a first step toward the application of MAS to space–time modulated metasurfaces.
\end{abstract}

%


%

%
\IEEEpeerreviewmaketitle

\section{Introduction}
\label{sec:Introduction}
\IEEEPARstart{E}{lectromagnetic} (EM) metasurfaces have attracted significant attention due to their ability to control waves using electrically thin arrangements of subwavelength resonators \cite{achouri2021electromagnetic,zahra2021electromagnetic}. By properly designing the unit cells, metasurfaces can realize a variety of functionalities such as beam steering \cite{naqvi2019beam, ashraf2023intelligent}, focusing \cite{wu2021design, li2021optically}, anomalous reflection and refraction \cite{liu2022review}, illusions \cite{smy2020surface}, and radiation pattern control \cite{hassan2024efficient}. The analysis of such structures using full-wave simulations is often computationally expensive, since the small geometrical details of the unit cells must be resolved over electrically large domains. For this reason, metasurfaces are commonly modeled as zero-thickness sheets described by surface susceptibility tensors and governed by the Generalized Sheet Transition Conditions (GSTCs) \cite{achouri2015general, ghaneizadeh2024generalized, holloway2016homogenization}. This approach replaces the physical thickness of the metasurface with equivalent boundary conditions that relate the discontinuities of the fields to induced surface polarizations, leading to more efficient numerical treatments. The GSTC-based model is applicable when the thickness is electrically small compared to the wavelength and when the unit cells are sufficiently subwavelength so that the metasurface can be homogenized. Under these conditions, the macroscopic scattering behavior is accurately described without explicitly modeling the volumetric structure.

In most GSTC-based formulations, metasurfaces are assumed to be spatially non-dispersive, meaning that the surface susceptibilities are considered independent of the incidence angle. This assumption is valid when the unit cells are deeply sub-wavelength and the EM response is purely local. However, in many practical metasurfaces, the unit-cell dimensions are not negligible compared to the wavelength, and the induced surface currents depend not only on the local fields but also on the fields in neighboring regions. In such cases, spatial dispersion (SD) is inherently present and may significantly affect the scattering behavior of the structure. Therefore, taking SD into account is essential for accurately modeling realistic metasurfaces. To address this limitation, extended GSTC formulations for spatially dispersive metasurfaces have been developed in 
\cite{rahmeier2023zero, gupta_part2, dugan2023spatially, dugan2023field, dugan2024surface}, where the metasurface response is represented in the spatial frequency domain, and the angular dependence of the surface susceptibilities is modeled as a ratio of two polynomials of the transverse wavenumber. 
This representation leads to extended GSTCs including spatial derivatives of the field differences and the average fields on the metasurface. 
These conditions were then incorporated in an integral-equation solver based on the Boundary Element Method (BEM), resulting in the IE-GSTC-SD framework for analyzing spatially dispersive metasurfaces.

The objective of this paper is to integrate the extended GSTC formulation for spatially dispersive metasurfaces into the Method of Auxiliary Sources (MAS) \cite{papakanellos2024method,kouroublakis2026analysis}, also known as Source Model Technique (SMT) \cite{leviatan1988generalized}, Method of Fundamental Solutions (MFS) ~\cite{ Karageorghis2025} and Multifilament Current Model (MFCM) (under the latter name it was successfully used to simulate spatially non-dispersive metasurfaces \cite{wang2020simulation}). While the IE-GSTC-SD framework based on the BEM provides an efficient tool for analyzing spatially dispersive structures, it still relies on surface discretization and meshing of the metasurface contour. In contrast, MAS is a meshless technique in which the EM fields are represented using discrete auxiliary sources placed outside the solution domain. This feature makes MAS simple to implement, flexible for handling different geometries, and computationally efficient. For these reasons, the integration of spatially dispersive GSTCs into the MAS framework provides an attractive alternative for the analysis of both infinite and finite metasurfaces. In addition, the developed method serves as a first step toward the application of MAS to more general space-time modulated metasurfaces. In a previous work, we developed a time-domain MAS formulation for frequency-dispersive metasurfaces without SD
\cite{kouroublakis2026timedomain}. Extending the method to include SD in the frequency domain is a necessary intermediate step before addressing fully space–time modulated metasurfaces \cite{wang2022pseudorandom, tiukuvaara2021floquet}. The formulation presented in this paper establishes this foundation by incorporating the extended GSTCs into MAS and validating the approach for a variety of spatially dispersive metasurface geometries. In what follows, we refer to this formulation as MAS-SD.  

The remainder of this paper is organized as follows. Section II presents the theoretical background of spatially dispersive metasurfaces and summarizes the extended GSTC formulation, with emphasis on the works of Gupta and co-workers \cite{rahmeier2023zero, gupta_part2, dugan2023spatially, dugan2023field, dugan2024surface}. In Section III, MAS is adapted to incorporate the extended GSTCs, and the MAS-SD formalism is developed for different geometrical configurations. Section IV presents numerical results that demonstrate the validity and accuracy of the proposed approach for both infinite and finite metasurface structures. Finally, the paper is concluded with a summary of the main findings and concluding remarks.  

An $\exp( j \omega t)$ time dependence is assumed and suppressed throughout.

\section{Theory of Spatially Dispersive Metasurfaces}
\label{sec:section_II}
\subsection{Spatially Non-Dispersive Metasurfaces}

\begin{figure} [htb!]
\centering
\includegraphics[scale=0.26]{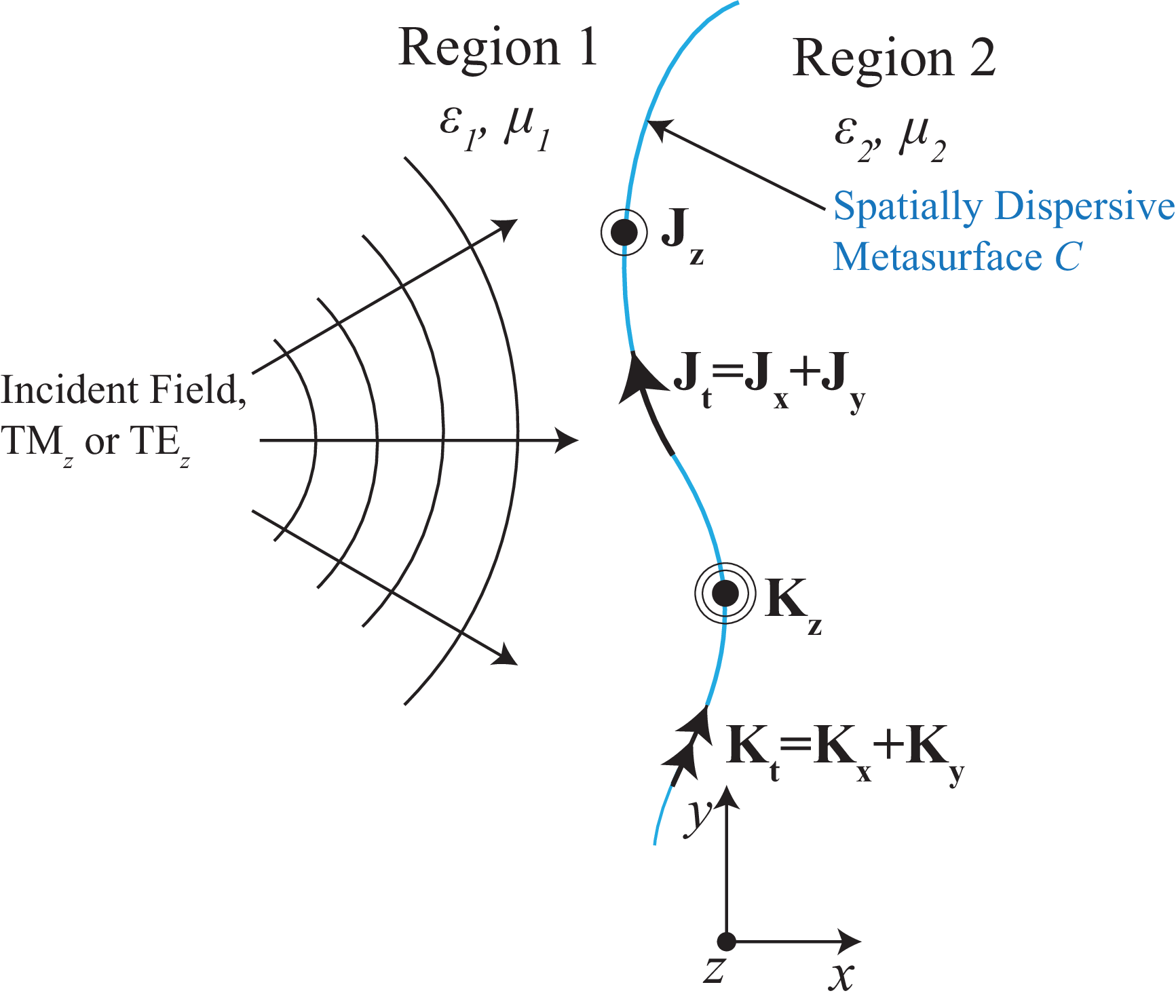}
\caption{A spatially dispersive metasurface is illuminated by an incident wave.}
\label{fig:Arbitrary_original}
\end{figure}

Consider a zero-thickness metasurface of arbitrary shape extending to infinity along the \(z\)-axis, as illustrated in Fig. \ref{fig:Arbitrary_original}. The metasurface separates the entire space into Region 1 (\(R_1\)) and Region 2 (\(R_2\)). A $\rm TM_z$ incident wave (i.e., with its electric field parallel to the \(z\)-axis) originating from \(R_1\) impinges on the metasurface generating scattered fields in both regions. 

The response of the metasurface to the incident field is associated with the development of macroscopic electric $\mathbf P$ and magnetic $\mathbf M$ polarizations on its surface. Consequentially, the scattered fields originate from the effective electric and magnetic surface currents associated with the induced polarizations of the metasurface, which microscopically arise from a combination of conduction and displacement currents within the meta-cells, comprising the metasurface. In the general case, these current densities 
include both $z$-directed and transverse tangential electric components $\mathbf{J}_z$, $\mathbf{J}_t$, as well as magnetic components $\mathbf{K}_z$, $\mathbf{K}_t$. As a result, the scattered fields consist of both $\rm TM_z$ and $\rm TE_z$ polarizations, regardless of the type of the incident wave.

For computing the total electric and magnetic fields in the two regions, we assume that the effect of the metasurface can be expressed through the GSTCs, which relate the field jumps across the metasurface to the macroscopic electric and magnetic polarizations of the metasurface, and are given by \cite{rahmeier2023zero}
\begin{equation}
\Delta \mathbf{H}
=
 -j \omega \hat{\mathbf{n}} \times \mathbf{P}_{\parallel}
-
 \nabla_{\parallel} \left(\frac{M_n}{\mu} \right)
\label{eq:gstc1}
\end{equation}
\begin{equation}
\Delta \mathbf{E}
=
  j \omega \mu_0 \hat{\mathbf{n}} \times \mathbf{M}_{\parallel}
-
 \nabla_{\parallel}
\left(
\frac{P_n}{\varepsilon}
\right),
\label{eq:gstc2}
\end{equation}
where $\varepsilon,~\mu$ are the constitutive parameters of the metasurface, $\Delta \mathbf{F} = \mathbf{F}^{+} - \mathbf{F}^{-}$ ($\mathbf F \in \mathbf E,\mathbf H$) denotes the discontinuity of the tangential field components across the metasurface, 
$\hat {\mathbf n}$ the unit normal vector at each point of the metasurface pointing towards $R_2$, and the subscripts $(\cdot)_{\parallel}$ and $(\cdot)_n$ denote tangential and normal components with respect to the surface while the operator $\nabla_{\parallel}$ represents the surface (transverse) gradient. From this point onward, we neglect the terms of the polarizations in the direction of $\hat {\mathbf n}$, i.e., normal to the metasurface. This is justified when the thickness of the metasurface is sufficiently smaller than the operating wavelength.

Since one side of the metasurface faces $R_1$ and the other faces $R_2$, its induced electric and magnetic polarizations are related, via the appropriate susceptibility tensors, to the average of field values on its two sides. Assuming a local (spatially non-dispersive) response, these relations can be written in the most general linear bianisotropic form as
\begin{equation}
\mathbf{P}
=
\varepsilon_0 \,\overline{\overline{\chi}}_{ee} \cdot \mathbf{E}_{\mathrm{av}}
+
\sqrt{\varepsilon_0\mu_0}\,\overline{\overline{\chi}}_{em} \cdot \mathbf{H}_{\mathrm{av}},
\label{eq:P_general}
\end{equation}
\begin{equation}
\mathbf{M}
=
\mu_0 \,\overline{\overline{\chi}}_{mm} \cdot \mathbf{H}_{\mathrm{av}}
+
\frac{1}{\eta_0}\,\overline{\overline{\chi}}_{me} \cdot \mathbf{E}_{\mathrm{av}},
\label{eq:M_general}
\end{equation}
where $\eta_0=\sqrt{\mu_0/\varepsilon_0}$ is the intrinsic impedance of free space and the average fields are defined by
\begin{equation}
\mathbf{F}_{\mathrm{av}}=\frac{1}{2}\left(\mathbf{F}^{+}+\mathbf{F}^{-}\right),
\quad \mathbf{F}\in\{\mathbf{E},\mathbf{H}\}.
\label{eq:average_fields}
\end{equation}
Here, $\overline{\overline{\chi}}_{ee}$ and $\overline{\overline{\chi}}_{mm}$ are the electric and magnetic susceptibility tensors, while $\overline{\overline{\chi}}_{em}$ and $\overline{\overline{\chi}}_{me}$ are the magnetoelectric coupling terms responsible for the bianisotropy. These $3\times 3$ tensors are expressed as
\begin{equation}
\overline{\overline{\chi}}_{ab}=\begin{bmatrix}
 \chi^{xx}_{\rm ab} & \chi^{xy}_{\rm ab} & \chi^{xz}_{\rm ab} \\
\chi^{yx}_{\rm ab} & \chi^{yy}_{\rm ab} & \chi^{yz}_{\rm ab} \\
\chi^{zx}_{\rm ab} & \chi^{zy}_{\rm ab} & \chi^{zz}_{\rm ab}
\end{bmatrix}
\end{equation}
with $ab \in \{ee, mm, em, me\}$. 

To make the tensorial coupling explicit, we write the polarization components in expanded form. For example, the $z$-directed electric surface polarization of (\ref{eq:P_general}) takes the form
\begin{multline}
P_z
=
\varepsilon_0\left(
\chi^{zx}_{ee} E_{x,\mathrm{av}}
+
\chi^{zy}_{ee} E_{y,\mathrm{av}}
+\chi^{zz}_{ee}E_{z,\rm av}\right)
+\\
\sqrt{\varepsilon_0\mu_0}\left(
\chi^{zx}_{em} H_{x,\mathrm{av}}
+
\chi^{zy}_{em} H_{y,\mathrm{av}}
+\chi^{zz}_{em} H_{z,\mathrm{av}}\right).
\label{eq:Pz_expanded}
\end{multline}


\subsection{Spatially Dispersive Metasurfaces}
For spatially dispersive metasurfaces by definition, the induced surface polarizations at a given point are not determined solely by the fields at that same point, but rather by the fields over the entire surface. This non-local response is modeled by replacing the point-wise products in \eqref{eq:P_general}--\eqref{eq:M_general} with spatial convolutions taken along the metasurface, namely,
\begin{equation}
\mathbf{P}(\boldsymbol{\rho})
=
\varepsilon_0\,
\overline{\overline{\chi}}_{ee}(\boldsymbol{\rho}) \ast \mathbf{E}_{\mathrm{av}}(\boldsymbol{\rho})
+
\sqrt{\varepsilon_0\mu_0}\,
\overline{\overline{\chi}}_{em}(\boldsymbol{\rho}) \ast \mathbf{H}_{\mathrm{av}}(\boldsymbol{\rho}),
\label{eq:P_conv_general}
\end{equation}
\begin{equation}
\mathbf{M}(\boldsymbol{\rho})
=
\mu_0\,
\overline{\overline{\chi}}_{mm}(\boldsymbol{\rho}) \ast \mathbf{H}_{\mathrm{av}}(\boldsymbol{\rho})
+
\frac{1}{\eta_0}\,
\overline{\overline{\chi}}_{me}(\boldsymbol{\rho}) \ast \mathbf{E}_{\mathrm{av}}(\boldsymbol{\rho}),
\label{eq:M_conv_general}
\end{equation}
with $\boldsymbol{\rho}=x \hat{\mathbf x}+ y\,\hat{\mathbf{y}}$ a point on the metasurface and $\ast$ the convolution along its tangential coordinates. 
For example, the $z$-directed electric surface polarization becomes
%
\begin{multline}
P_z(\boldsymbol{\rho})
=
\varepsilon_0\left(
\chi^{zx}_{ee} \ast E_{x,\mathrm{av}}
+
\chi^{zy}_{ee} \ast E_{y,\mathrm{av}}
+
\chi^{zz}_{ee} \ast E_{z,\mathrm{av}}
\right)
+\\
\sqrt{\varepsilon_0\mu_0}\left(
\chi^{zx}_{em} \ast H_{x,\mathrm{av}}
+
\chi^{zy}_{em} \ast H_{y,\mathrm{av}}
+
\chi^{zz}_{em} \ast H_{z,\mathrm{av}}
\right).
\label{eq:Pz_conv}
\end{multline}

\begin{figure}[htb!]
\centering
\includegraphics[width=0.35\textwidth]{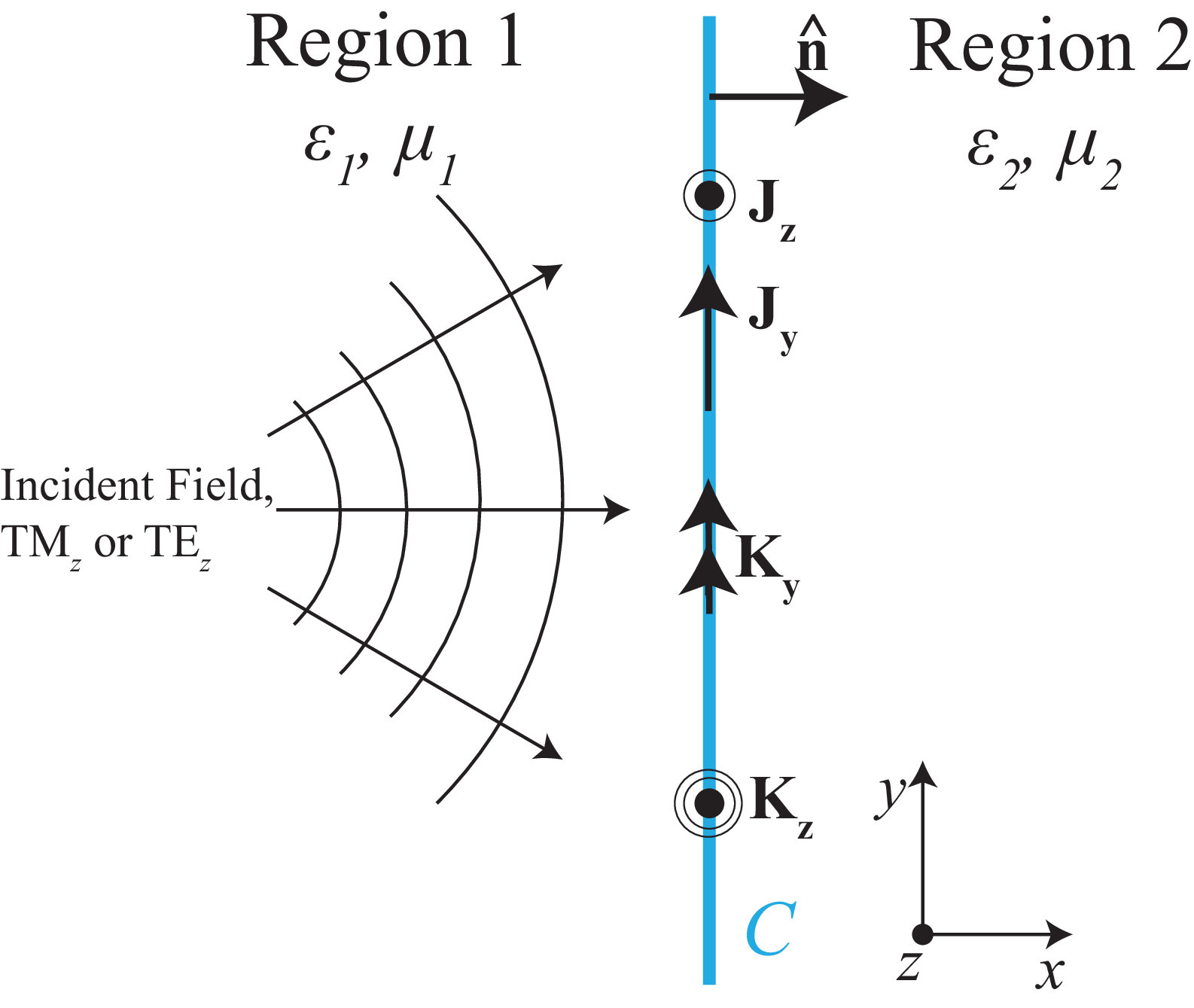}    
\caption{A spatially dispersive planar metasurface in the $yz$ plane.}
\label{fig:infinite_planar_intro}
\end{figure}

\paragraph{Application to Planar Metasurfaces}
Consider the planar metasurface of Fig.~\ref{fig:infinite_planar_intro} lying in the \(yz\)-plane (hence \(\hat{\mathbf n}=\hat{\mathbf x}\)).

We assume that SD occurs only along the \(y\)-axis (transverse tangential direction). Below, we will neglect the bianisotropic terms, which essentially means that the metasurface is simply anisotropic (mixes the polarizations of the scattered fields symmetrically with respect to the incidence direction). It should be emphasized, that this simplification does not entail any loss of generality regarding the application of the method to spatially dispersive metasurfaces. 
From \eqref{eq:gstc1} and \eqref{eq:gstc2}, we obtain 
\begin{subequations}
\label{eq:Diff_Pol}
\begin{equation}
\Delta H_y =  j\omega P_z, 
\qquad 
\Delta H_z = - j\omega  P_y,
\label{eq:DH_from_P}
\end{equation}
\begin{equation}
\Delta E_z =  j\omega M_y,
\qquad
\Delta E_y = - j\omega M_z.
\label{eq:DE_from_M}
\end{equation}
\end{subequations}
\noindent 

The convolution equations (\ref{eq:Pz_conv}) are expressed in the spatial frequency domain of the transverse wavevector $k_y$ using the Fourier transform $\mathcal{F}_t(\cdot)$ yielding
\cite{rahmeier2023zero,gupta_part2}
\begin{subequations}
\label{eq:spat_freq_polar}
\begin{equation}
\tilde{\mathbf{P}}(k_y)
=
\varepsilon_0\,\overline{\overline{\chi}}_{ee}(k_y)\cdot \tilde{\mathbf{E}}_{\mathrm{av}}(k_y)
,
\end{equation}
\begin{equation}
\tilde{\mathbf{M}}(k_y)
=
\mu_0\,\overline{\overline{\chi}}_{mm}(k_y)\cdot \tilde{\mathbf{H}}_{\mathrm{av}}(k_y).
\end{equation}
\end{subequations}
which are simple products of the average fields and the surface
susceptibilities, as opposed to convolution in space, making it
suitable for compact unit cell description and subsequent
numerical computation. Substituting ~(\ref{eq:spat_freq_polar}) into ~(\ref{eq:Diff_Pol}), we get, for example,
\begin{equation}
\label{eq:spatial_freq_dom}
\Delta H_y=  j \omega \varepsilon_0\chi^{zy}_{ee}(k_y)E_{y, \rm av}(k_y)+ j \omega \varepsilon_0\chi^{zz}_{ee}(k_y)E_{z, \rm av}(k_y)
\end{equation}
\noindent and similar expressions for $\Delta E_z$, $\Delta H_z$, and $\Delta E_y$, which are omitted for brevity.

In the spatial frequency domain, each susceptibility component can be represented by a rational function of the transverse wavenumber $k_y$ \cite{rahmeier2023zero}. For a generic tensor entry $\chi^{\alpha\beta}_{ab}(k_y)$, with $ab\in\{ee,mm,em,me\}$ and $\alpha,\beta\in\{y,z\}$, we have
\begin{equation}
\chi^{\alpha\beta}_{ab}(k_y)
=
\frac{\displaystyle \sum_{q=0}^{Q} a^{\alpha\beta}_{ab,q}\,k_y^{q}}
{\displaystyle \sum_{n=0}^{N} b^{\alpha\beta}_{ab,n}\,k_y^{n}}.
\label{eq:rational_recall}
\end{equation}
This representation accounts for the zeros and poles of the surface susceptibilities, enabling an accurate description of the metasurface response for varying angles of plane-wave incidence. With this representation, (\ref{eq:spatial_freq_dom}) is written as
\begin{subequations}
\label{eq:ratio_pol}
\begin{equation}
\Delta H^y_y=  j \omega \varepsilon_0\frac{\sum_{q=0}^{Q}a^{zy}_{ee,q}k^q_y}{\sum_{n=0}^{N}b^{zy}_{ee,n}k^n_y}E_{y, \rm av}(k_y)
\end{equation}
\begin{equation}
\Delta H^z_y=  j \omega \varepsilon_0\frac{\sum_{q=0}^{Q}a^{zz}_{ee,q}k^q_y}{\sum_{n=0}^{N}b^{zz}_{ee,n}k^n_y}E_{z, \rm av}(k_y)
\end{equation}
\end{subequations}
obtaining a polynomial relation between field jumps and average fields in $k_y$. Here, $\Delta H^{y}_y$ ($\Delta H^z_y$) is the portion of $\Delta H_y$ related to $E_{y,av}$ ($E_{z,av}$). Similar equations can be obtained for the portions of $\Delta E_z$, $\Delta H_z$, and $\Delta E_y$. In ~(\ref{eq:ratio_pol}), multiplying by the common denominator and transforming back to the spatial domain yields
\begin{subequations}
\label{eq:extended_GSTC}
\begin{multline}
\sum_{n=0}^{N}b^{zy}_{ee,n} j^n\frac{\partial^n }{\partial y^n}\Delta H^y_y(y)=
 j \omega \varepsilon_0\sum_{q=0}^{Q}a^{zy}_{ee,q} j^q\frac{\partial^q }{\partial y^q}E_{y, \rm av}(y)
\end{multline}
\begin{multline}
\sum_{n=0}^{N}b^{zz}_{ee,n} j^n\frac{\partial^n }{\partial y^n}\Delta H^z_y(y)=
 j \omega \varepsilon_0\sum_{q=0}^{Q}a^{zz}_{ee,q} j^q\frac{\partial^q }{\partial y^q}E_{z, \rm av}(y)
\end{multline}
\end{subequations}

Equations \eqref{eq:extended_GSTC}, along with those for the portions of $\Delta E_z$, $\Delta H_z$, and $\Delta E_y$, represent the  \emph{extended GSTCs} for the anisotropic planar metasurface. The coefficient sets $\{a^{\alpha\beta}_{ab,m}\}$ and $\{b^{\alpha\beta}_{ab,n}\}$ are obtained by fitting to unit-cell simulation data over incidence angle (equivalently over $k_y$). The corresponding equations of \eqref{eq:ratio_pol} for bianisotropic metasurfaces contain two terms on the right-hand side instead of one, with each term involving a rational polynomial expression with different denominators (the additional term corresponds to the magnetoelectric susceptibilities). Thus, the Fourier transform of these equations into the spatial domain, leading to the extended GSTCs for the bianisotropic case, is not as straightforward as for the anisotropic metasurface and will not be shown in this work. 

\begin{figure}[htb!]
\centering
\subfigure[]{\includegraphics[width=0.35\textwidth]{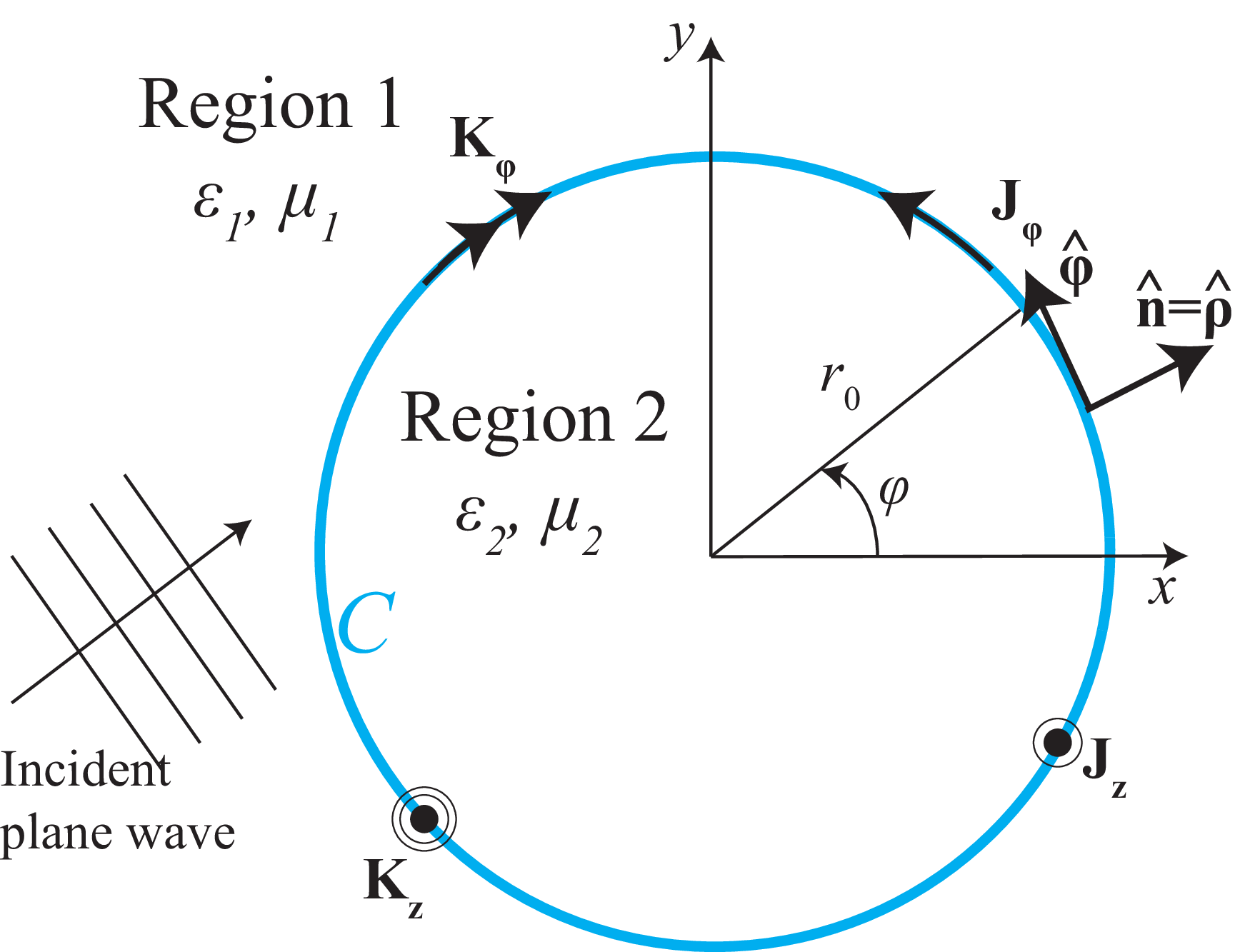}}    
\caption{A spatially dispersive circular cylindrical metasurface.}
\label{fig:closed_cylinder_intro}
\end{figure}

\paragraph{Application to Cylindrical Metasurfaces}
Consider the circular cylindrical metasurface of radius \(r_0\) 
depicted in Fig. \ref{fig:closed_cylinder_intro}. Here, the susceptibility tensors must be defined locally for every point on the zero-thickness metasurface. Hence, at each point of $C$, we define a local coordinate system which, due to the circular boundary, coincides with the global cylindrical coordinate system. For this case, \eqref{eq:spat_freq_polar} should be replaced by
%

\begin{subequations}
\begin{equation}
\tilde{\mathbf{P}}(k_{\phi})
=
\varepsilon_0\,\overline{\overline{\chi}}_{ee}(k_{\phi})\cdot \tilde{\mathbf{E}}_{\mathrm{av}}(k_{\phi}),
\end{equation}
\begin{equation}
\tilde{\mathbf{M}}(k_{\phi})
=
\mu_0\,\overline{\overline{\chi}}_{mm}(k_{\phi})\cdot \tilde{\mathbf{H}}_{\mathrm{av}}(k_{\phi}),
\end{equation}
\end{subequations}
where \(k_{\phi}\) is the azimuthal wavevector. 
The relations between the field discontinuities and the EM polarizations for this geometry are obtained applying  \eqref{eq:gstc1} and \eqref{eq:gstc2} for $\hat {\mathbf n}= \hat{\boldsymbol{\rho}}$, yielding
\begin{subequations}
\begin{equation}
\Delta H_\phi =  j\omega P_z, 
\qquad 
\Delta H_z = - j\omega P_\phi,
\end{equation}
\begin{equation}
\Delta E_z =  j\omega M_\phi,
\qquad
\Delta E_\phi = - j\omega M_z.
\end{equation}
\end{subequations}
In the spatial frequency domain, the susceptibilities are given by ~(\ref{eq:rational_recall}) with $k_\phi$ in place of $k_y$.
%
%
The corresponding GSTC equations are obtained by replacing $\partial y=r_0 \partial\phi$ in \eqref{eq:extended_GSTC}, i.e.,
\begin{subequations}
\label{eq:ext_GSTC_cyl}
\begin{multline}
\sum_{n=0}^{N}b^{z \phi}_{ee,n} j^n\frac{\partial^n }{r_0^n\partial \phi^n}\Delta H^\phi_\phi(\phi)= \\ j \omega \varepsilon_0\sum_{q=0}^{Q}a^{z \phi}_{ee,q} j^q\frac{\partial^q }{r_0^q\partial \phi^q}E_{\phi, \rm av}(\phi)
\end{multline}
\begin{multline}
\sum_{n=0}^{N}b^{zz}_{ee,n} j^n\frac{\partial^n }{r_0^n\partial \phi^n}\Delta H^z_\phi(\phi)= \\ j \omega \varepsilon_0\sum_{q=0}^{Q}a^{z z}_{ee,q} j^q\frac{\partial^q }{r_0^q\partial \phi^q}E_{z, \rm av}(\phi)
\end{multline}
Similar equations are derived for the portions of $\Delta E_z$, $\Delta H_z$,  $\Delta H_\phi$ completing the extended GSTCs for the anisotropic case.

\end{subequations}

\subsection{Lorentz-Type Spatial Dispersion Model}
\label{sec:Lorentzian_res}
A physically meaningful and practically relevant specialization of the rational-polynomial susceptibility model is obtained by assuming that the metasurface response is described by Lorentz-type resonators \cite{rahmeier2023zero}. This is an isotropic metasurface, i.e., the scattered fields have the same polarization as the incident field. 
For a given electric polarization component, e.g., $P_z$, the frequency-domain Lorentz oscillator model is
\begin{equation}
-\omega^2 \tilde{P}_z
+
 j\gamma \omega \tilde{P}_z
+
\omega_0^2 \tilde{P}_z
=
\varepsilon_0 \omega_p^2 \tilde{E}_{z,\mathrm{av}},
\label{eq:lorentz_basic}
\end{equation}
with $\omega_0$ the resonance frequency, $\omega_p$ the plasma frequency, and $\gamma$ the damping coefficient. In planar spatially dispersive metasurfaces, these parameters depend on the transverse wavenumber $k_y$. Expanding them around normal incidence ($k_y=0$) and retaining terms up to second order, yields
\begin{subequations}
\label{eq:taylor_expansion}
\begin{equation}
\omega_0^2(k_y) \approx \zeta_0^2 +\zeta_1k_y+ \zeta_2 k_y^2,
\end{equation}
\begin{equation}
\omega_p^2(k_y) \approx \beta_0^2 +\beta_1k_y+ \beta_2 k_y^2,
\end{equation}
\begin{equation}
\gamma(k_y) \approx \alpha_0 + \alpha_1 k_y+ \alpha_2 k_y^2.
\end{equation}
\end{subequations}
Substituting (\eqref{eq:taylor_expansion}) into  \eqref{eq:lorentz_basic} enables the relation between the average fields and the polarization to be written as
\begin{equation}
\tilde{\mathcal P}^z_z=\varepsilon_0\frac{\chi^{zz}_{ee,0}+ j \chi^{zz}_{ee,1}k_y-\chi^{zz}_{ee,2}k_y^2}{1+ j\xi^{zz}_{ee,1}k_y-\xi^{zz}_{ee,2}k_y^2}\tilde{\mathcal E}_{z,\mathrm {av}}
\label{eq:lorentzian_pol}
\end{equation}
where
\begin{subequations}
\label{eq:Lorentzian_par}
\begin{equation}
\xi^{zz}_{ee,1}=\frac{\omega\alpha_1- j \zeta_1}{\zeta_0^2-\omega^2+ j \omega\alpha_0}
\end{equation}
\begin{equation}
\xi^{zz}_{ee,2}=\frac{-(\zeta_2+ j \omega\alpha_2)}{\zeta_0^2-\omega^2+ j\omega\alpha_0}
\end{equation}
\begin{equation}
\chi^{zz}_{ee,0}=\frac{\beta_0^2}{\zeta_0^2-\omega^2+ j \omega\alpha_0}
\end{equation}
\begin{equation}
\chi^{zz}_{ee,1}=\frac{ j\beta_1}{\zeta_0^2-\omega^2+ j \omega\alpha_0}
\end{equation}
\begin{equation}
\chi^{zz}_{ee,2}=\frac{-\beta_2}{\zeta_0^2-\omega^2+ j \omega\alpha_0}
\end{equation}
\end{subequations}
For spatially symmetric unit cells, the terms $\xi_1$ in (\ref{eq:lorentzian_pol}) are zero. By the inverse spatial Fourier transform $\mathcal F_y^{-1}$ in (\ref{eq:lorentzian_pol}), we get 
\begin{subequations}
\label{eq:GSTC_pl_lorentzian}
\begin{equation}
\left[ \xi^{zz}_{ee,2}\frac{\partial^2 }{\partial y^2}+1\right]\Delta H_y= j\omega \varepsilon_0\left[ \chi^{zz}_{ee,2}\frac{\partial^2 }{\partial y^2} + \chi^{zz}_{ee,0}\right]{E}_{z,\rm av},
\end{equation}
\begin{multline}
\left[ \xi^{yy}_{mm,2}\frac{\partial^2 }{\partial y^2}+1\right]\Delta E_z=\\ j\omega \mu_0\left[ \chi^{yy}_{mm,2}\frac{\partial^2 }{\partial y^2} + \chi^{yy}_{mm,0}\right]{H}_{y,\rm av},
\end{multline}
\end{subequations}
which give the complete set of GSTCs for this isotropic case. For cylindrical metasurfaces, the corresponding GSTCs are
\begin{subequations}
\label{eq:GSTC_cylindrical}
\begin{equation}
\left[ \xi^{zz}_{ee,2}\frac{\partial^2 }{r_0^2\partial \phi^2}+1\right]\Delta H_\phi= j\omega\varepsilon_0\left[ \chi^{zz}_{ee,2}\frac{\partial^2 }{r_0^2\partial \phi^2} + \chi^{zz}_{ee,0}\right]{E}_{z,\rm av},
\end{equation}
\begin{multline}
\left[ \xi^{\phi \phi}_{mm,2}\frac{\partial^2 }{r_0^2\partial \phi^2}+1\right]\Delta E_z=\\ j\omega \mu_0\left[ \chi^{\phi \phi}_{mm,2}\frac{\partial^2 }{r_0^2\partial \phi^2} + \chi^{\phi \phi}_{mm,0}\right]{H}_{\phi,\rm av}.
\end{multline}
\end{subequations}






\section{MAS-SD Implementation}
In this section, we present the application of MAS-SD for the implementation of the extended GSTCs introduced in Section \ref{sec:section_II}. Specifically, we demonstrate how MAS-SD is applied to the following spatially dispersive metasurfaces:  
1) Infinitely extended planar metasurface,  
2) Finite planar metasurface,   
3) Semicircular cylindrical metasurface,  
4) Circular cylindrical metasurface, and
5) Open polygonal cylindrical metasurface. 

The MAS-SD formulation is presented for the general case of spatially dispersive metasurfaces. For clarity, emphasis is placed on metasurfaces described by Lorentzian resonators, since the numerical examples considered in this work correspond to this simpler isotropic configuration.
This choice is made without loss of generality, as the extension to anisotropic and bianisotropic metasurfaces follows directly from the same formulation. As will be shown below, each geometrical configuration requires a specific placement of the auxiliary sources, and, therefore, the different cases are examined separately.

\subsection {Infinite Planar Metasurface}
\label{sec:infinite_planar}
Consider an infinite planar metasurface on the \(yz\)-plane (see Fig.~\ref{fig:infinite_planar_intro}). This idealized geometry approximates the realistic case when the operating wavelength is much smaller than the dimensions of the metasurface (still large enough to assume zero thickness of the metasurface). 
The objective is to compute the transmitted fields in $R_2$ and the reflected fields in $R_1$ due to the induced surface currents on the metasurface \((\mathbf{J}_z,\mathbf{J}_y,\mathbf{K}_z,\mathbf{K}_y)\). 
\begin{figure}[htb!]
\centering
\subfigure[]{\includegraphics[width=0.35\textwidth]{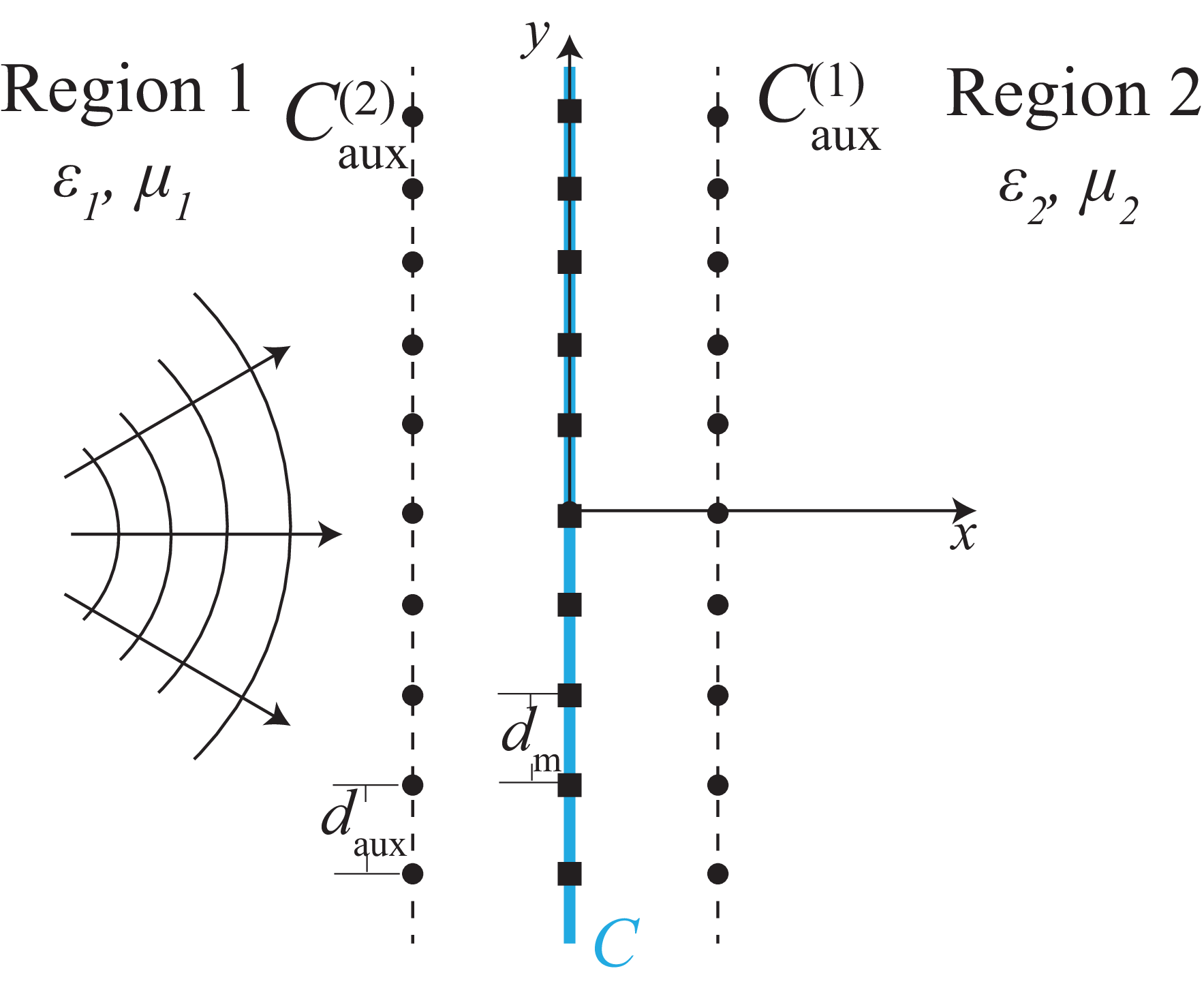}}    
\caption{Application of MAS-SD on an infinite planar metasurface excited by a gaussian beam wave. Black circles represent auxiliary sources and black squares matching points.}
\label{fig:infinite_MAS}
\end{figure}

In the application of MAS, we consider a problem equivalent to the original one (depicted in Fig.~\ref{fig:infinite_MAS}), in which the fields in \(R_1\) and \(R_2\) originate from current densities defined on two auxiliary surfaces, \(C^{(1)}_{\mathrm{aux}}\) and \(C^{(2)}_{\mathrm{aux}}\). Both auxiliary surfaces are parallel to the metasurface boundary $C$. 
Specifically, the reflected fields in $R_1$ are produced by $N_1$ collocated $z$-directed infinitely long electric and magnetic current filaments with amplitudes $I^{(1)}_{zl}$ and $K^{(1)}_{zl}$, respectively, placed on $C^{(1)}_{\mathrm{aux}}$ and radiating in an unbounded medium the same as that of $R_1$. Similarly, the transmitted fields in $R_2$ are generated by $N_2$ collocated $z$-directed electric and magnetic current filaments with amplitudes $I^{(2)}_{zl}$ and $K^{(2)}_{zl}$ located on $C^{(2)}_{\mathrm{aux}}$ and radiating in an unbounded medium the same as that of $R_2$.

The auxiliary sources are placed equidistantly with spacing $d_{\mathrm{aux}}$ and symmetrically with respect to the $x$-axis. 
The total fields in each region are then obtained as superpositions of the fields radiated by the corresponding auxiliary sources. Assuming vacuum in $R_1$ and $R_2$, the fields radiated in region $p$ ($p=1,2$) by the 
$n$-th auxiliary electric current filament are
\begin{subequations}
\label{eq:TMz_fields}
\begin{equation}
\mathbf{E}^{(p)}_{n}(x,y)=-\frac{k_0\eta_0}{4}I_{pn}H_0^{\left(2\right)}\left(k_0D_{pn}\right)\hat{\mathbf{z}},
\end{equation}
\begin{align}
\mathbf{H}^{(p)}_{n}\left(x,y\right)=&\frac{k_0}{4 j}I_{pn}\frac{\hat{\mathbf{x}}\left(y_{pn}-y\right)+\hat{\mathbf{y}}\left(x-x_{pn}\right)}{D_{pn}}H_1^{\left(2\right)}\left(k_0D_{pn}\right),
\end{align}
\end{subequations}
while by the 
$n$-th auxiliary magnetic current filament are
\begin{subequations}
\label{eq:TEZ_fields}
\begin{equation}
\textbf{H}^{(p)}_{n}(x,y)=-\frac{k_0}{4\eta_0}K_{pn}H_0^{\left(2\right)}\left(k_0D_{pn}\right)\hat{\mathbf{z}},
\end{equation}
\begin{align}
\mathbf{E}^{(p)}_{n}\left(x,y\right)=&\frac{k_0}{4 j}K_{pn}\frac{\hat{\mathbf{x}}\left(y_{pn}-y\right)+\hat{\mathbf{y}}\left(x-x_{pn}\right)}{D_{pn}}
H_1^{\left(2\right)}\left(k_0D_{pn}\right),
\end{align}
\end{subequations}
with $k_0$ the vacuum wavenumber, 
$I_{pj}$ and $K_{pj}$ the amplitudes of the electric and magnetic current filaments, 
$(x_{pj}, y_{pj})$ the positions of the filaments radiating in $R_p$, and $D_{pj}$ the distance between the $j$-th filament and the observation point. 

Since the numerical examples considered in this work involve metasurfaces with Lorentz-type resonators analyzed in Section~\ref{sec:Lorentzian_res}, which are isotropic, an incident $\rm TM_z$ wave requires only electric auxiliary sources to generate the reflected and transmitted fields. Conversely, for an incident $\rm TE_z$ wave, only magnetic auxiliary sources are required. Assuming $\rm TM_z$ incidence from $R_1$ and applying the extended GSTCs (\ref{eq:GSTC_pl_lorentzian}) at $M$ matching points on the metasurface, we obtain a linear system
\begin{equation}
[Z]\mathbf I=\mathbf V,
\label{eq:linear_system}
\end{equation}
with $\mathbf I=[\mathbf I_2~\mathbf I_1]^{T}$ the column vector of the unknown complex amplitudes of the auxiliary currents, and $\mathbf I_1$ ($\mathbf I_2$) the $N_1\times 1$ ($N_2\times 1$) sub-vector with elements the complex amplitudes $I_{1n}$ ($I_{2n})$, while  $\mathbf V$ is the excitation vector given by
\begin{multline}
\mathbf V=\\\begin{bmatrix}
(\xi^{zz}_{ee,2}\frac{\partial^2 }{\partial y^2}+1)\mathbf H^{inc}+\frac{ j\omega \varepsilon_0}{2}(\chi^{zz}_{ee,2}\frac{\partial^2 }{\partial y^2}+\chi^{zz}_{ee,0})\mathbf E^{inc}\\
(\xi^{yy}_{mm,2}\frac{\partial^2 }{\partial y^2}+1)\mathbf E^{inc}+\frac{ j\omega \mu_0}{2}(\chi^{yy}_{mm,2}\frac{\partial^2 }{\partial y^2}+\chi^{yy}_{mm,0})\mathbf H^{inc}
\end{bmatrix}
\end{multline}
where $\mathbf E^{inc}$ ($\mathbf H^{inc}$) is an $M\times 1$ vector with elements the values of the incident tangential electric (magnetic) field at the matching points. The impedance matrix $[Z]$ is given in (\ref{eq:impedance_matrix}) where $\mathbf E_z^{(p)}$ and $ \mathbf H_y^{(p)}$ ($p=1,2$) are $M\times N_p$ matrices consisting of the fields' components samples at
the $M$ matching points. The matching points are placed equidistantly along \(C\) with spacing \(d_m\), and symmetrically with respect to $x$. 
The solution of the system yields the unknown amplitudes of the auxiliary currents. Then, the fields in \(R_1\) and \(R_2\) are computed by ~(\ref{eq:TMz_fields}) and (\ref{eq:TEZ_fields}) and using the superposition principle.
\begin{figure*}[htb!]
\begin{align}
\label{eq:impedance_matrix}
[Z]=\begin{bmatrix}
(\xi^{zz}_{ee,2}\frac{\partial^2 }{\partial y^2}+1)\mathbf H^{(2)}_y-\frac{ j\omega \varepsilon_0}{2}(\chi^{zz}_{ee,2}\frac{\partial^2 }{\partial y^2}+\chi^{zz}_{ee,0})\mathbf E^{(2)}_z &  -(\xi^{zz}_{ee,2}\frac{\partial^2 }{\partial y^2}+1)\mathbf H^{(1)}_y-\frac{ j\omega \varepsilon_0}{2}(\chi^{zz}_{ee,2}\frac{\partial^2 }{\partial y^2}+\chi^{zz}_{ee,0})\mathbf E^{(1)}_z \\
(\xi^{yy}_{mm,2}\frac{\partial^2 }{\partial y^2}+1)\mathbf E^{(2)}_z-\frac{ j\omega \mu_0}{2}(\chi^{yy}_{mm,2}\frac{\partial^2 }{\partial y^2}+\chi^{yy}_{mm,0})\mathbf H^{(2)}_y &- (\xi^{yy}_{mm,2}\frac{\partial^2 }{\partial y^2}+1)\mathbf E^{(1)}_z-\frac{ j\omega \mu_0}{2}(\chi^{yy}_{mm,2}\frac{\partial^2 }{\partial y^2}+\chi^{yy}_{mm,0})\mathbf H^{(1)}_y
\end{bmatrix}
\end{align}
\end{figure*}

\subsection{Finite Planar Metasurface}
\label{sec:finite_planar}
\begin{figure}[htb!]
\centering
\includegraphics[width=0.36\textwidth]{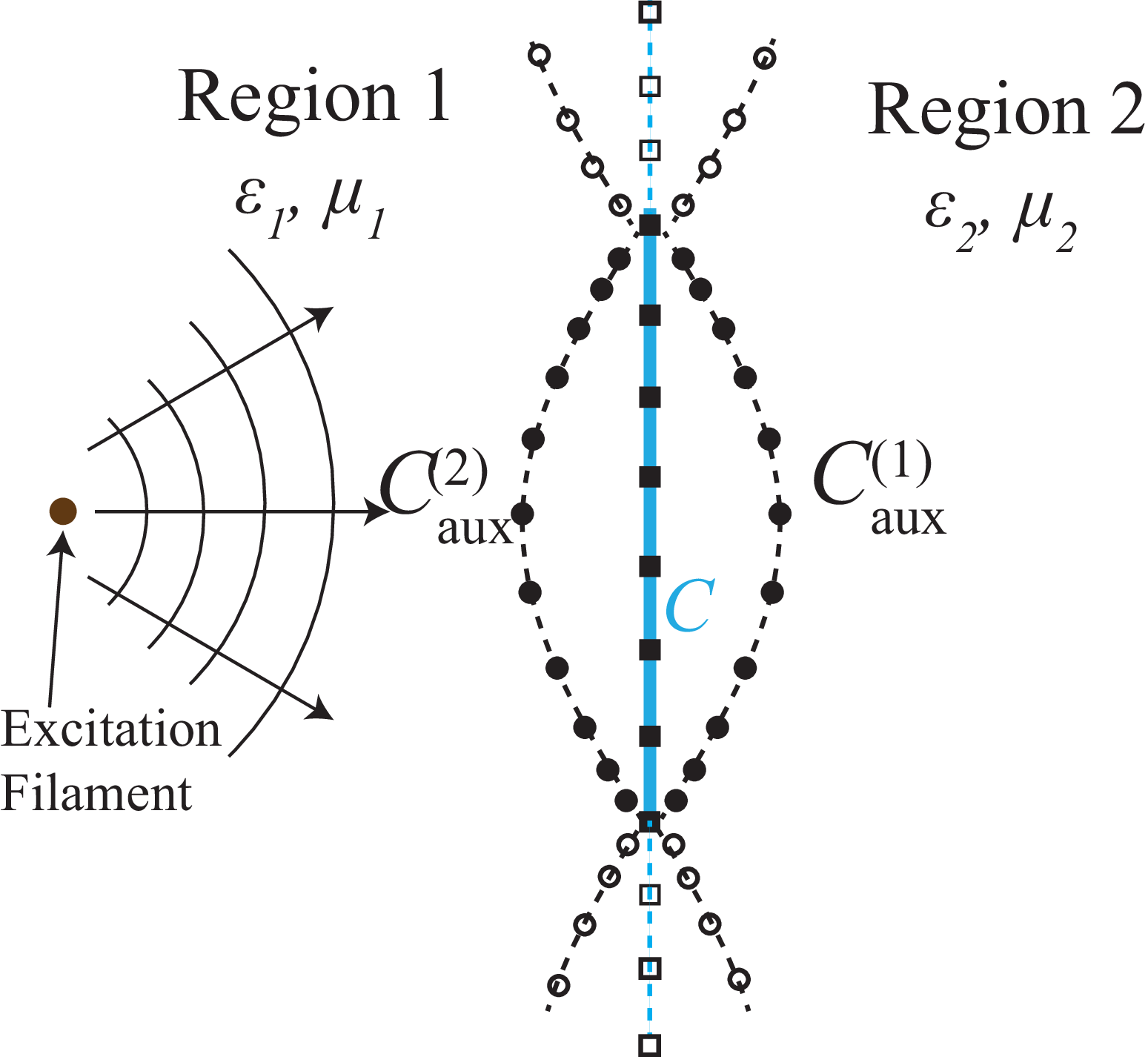}    
\caption{Application of MAS-SD on a finite planar metasurface excited by an electric current filament. Black circles represent auxiliary sources and black squares matching points. Blank circles and squares represent extended auxiliary sources and matching points.}
\label{fig:finite_MAS}
\end{figure}

Next, we analyze a finite planar metasurface. Essentially, it differs from the previous case in that the metasurface is finite along \(y\). The application of MAS-SD is shown in Fig.~\ref{fig:finite_MAS}. Due to the finite dimensions of the metasurface and the presence of edges, a different placement of the auxiliary sources is required. To better approximate the singular field behavior near the edges, sources are positioned closer to the edges and gradually placed farther away toward the interior. The locations of the $N_{m1}$ and $N_{m2}$ auxiliary sources opposite to the metasurface, shown as black circles in Fig.~\ref{fig:finite_MAS}(b), are selected according to the empirical rules of \cite{leviatan1991analysis, wang2020two}. In addition, $N_{e1}$ and $N_{e2}$ collocated electric and magnetic sources are placed above and below the metasurface, indicated by blank circles in Fig.~\ref{fig:finite_MAS}(b). 
This results in a total of $N_1 = N_{m1} + N_{e1}$ sources on $C^{(1)}_{\rm aux}$ and $N_2 = N_{m2} + N_{e2}$ sources on $C^{(2)}_{\rm aux}$.  Again, $R_1$ is located to the left of the metasurface and $R_2$ to the right. Contrary to the infinite metasurface of Section \ref{sec:infinite_planar}, the incident field now exists in both regions.


 
Considering the simple case of a metasurface with Lorentzian resonators, only electric or magnetic current filaments are used as auxiliary sources, depending on the incident polarization. Assuming $\mathrm{TM}_z$ incidence, the boundary condition \eqref{eq:GSTC_pl_lorentzian} is enforced at the $M_m$ matching points shown as black squares in Fig.~\ref{fig:finite_MAS}. These matching points are placed equidistantly along the metasurface, avoiding however the edge (singular) points. The tangential electric and magnetic field continuity conditions are enforced at the $M_e$ matching points shown as blank squares in Fig.~\ref{fig:finite_MAS}, located beyond the metasurface edges and distributed with the same spacing as the $M_m$ points. This procedure leads to a linear system of the form \eqref{eq:linear_system}, with 
excitation vector given by
\begin{align}
\mathbf V=\begin{bmatrix}
(\xi^{zz}_{ee,2}\frac{\partial^2 }{\partial y^2}+1)\mathbf H^{inc}+\frac{ j\omega \varepsilon_0}{2}(\chi^{zz}_{ee,2}\frac{\partial^2 }{\partial y^2}+\chi^{zz}_{ee,0})\mathbf E^{inc}\\
(\xi^{yy}_{mm,2}\frac{\partial^2 }{\partial y^2}+1)\mathbf E^{inc}+\frac{ j\omega \mu_0}{2}(\chi^{yy}_{mm,2}\frac{\partial^2 }{\partial y^2}+\chi^{yy}_{mm,0})\mathbf H^{inc}\\
\mathbf 0
\\
\mathbf 0 
\end{bmatrix}
\end{align}
where $\mathbf E^{inc}$ ($\mathbf H^{inc}$) is an $M_m\times 1$ vector, and $\mathbf 0$ corresponds to $M_e\times 1$ zero vector. The impedance matrix $[Z]$ is given by (\ref{eq:impedance_matrix_finite}) with first two rows corresponding to the GSTC condition applied at the $M_m$ matching points and the last two, to the continuity equations of the tangential fields at the $M_e$ matching points. 
Thus, \(\mathbf E^{pm}_z\) and \(\mathbf H^{pm}_y\) are \(M_m \times N_p\) matrices, whereas \(\mathbf E^{pe}_z\) and \(\mathbf H^{pe}_y\) are \(M_e \times N_p\) matrices. After solving the system, the fields in \(R_1\) and \(R_2\) follow via (\ref{eq:TMz_fields})–(\ref{eq:TEZ_fields}) and superposition.
\begin{figure*} [t!]
\begin{align}
\label{eq:impedance_matrix_finite}
\mathbf [Z]=\begin{bmatrix}
(\xi^{zz}_{ee,2}\frac{\partial^2 }{\partial y^2}+1)\mathbf H^{(2m)}_y-\frac{ j\omega \varepsilon_0}{2}(\chi^{zz}_{ee,2}\frac{\partial^2 }{\partial y^2}+\chi^{zz}_{ee,0})\mathbf E^{(2m)}_z &  -(\xi^{zz}_{ee,2}\frac{\partial^2 }{\partial y^2}+1)\mathbf H^{(1m)}_y-\frac{ j\omega \varepsilon_0}{2}(\chi^{zz}_{ee,2}\frac{\partial^2 }{\partial y^2}+\chi^{zz}_{ee,0})\mathbf E^{(1m)}_z \\
(\xi^{yy}_{mm,2}\frac{\partial^2 }{\partial y^2}+1)\mathbf E^{(2m)}_z-\frac{ j\omega \mu_0}{2}(\chi^{yy}_{mm,2}\frac{\partial^2 }{\partial y^2}+\chi^{yy}_{mm,0})\mathbf H^{(2m)}_y & -(\xi^{yy}_{mm,2}\frac{\partial^2 }{\partial y^2}+1)\mathbf E^{(1m)}_z-\frac{ j\omega \mu_0}{2}(\chi^{yy}_{mm,2}\frac{\partial^2 }{\partial y^2}+\chi^{yy}_{mm,0})\mathbf H^{(1m)}_y\\
\mathbf E^{(2e)}_z & -\mathbf E^{(1e)}_z\\
\mathbf H^{(2e)}_y & -\mathbf H^{(1e)}_y
\end{bmatrix}
\end{align}
\end{figure*}



\subsection{Open and Closed Circular Cylindrical Metasurfaces}
\label{sec:cylindrical}
\begin{figure}[htb!]
\centering
\subfigure[]{\includegraphics[width=0.20\textwidth]{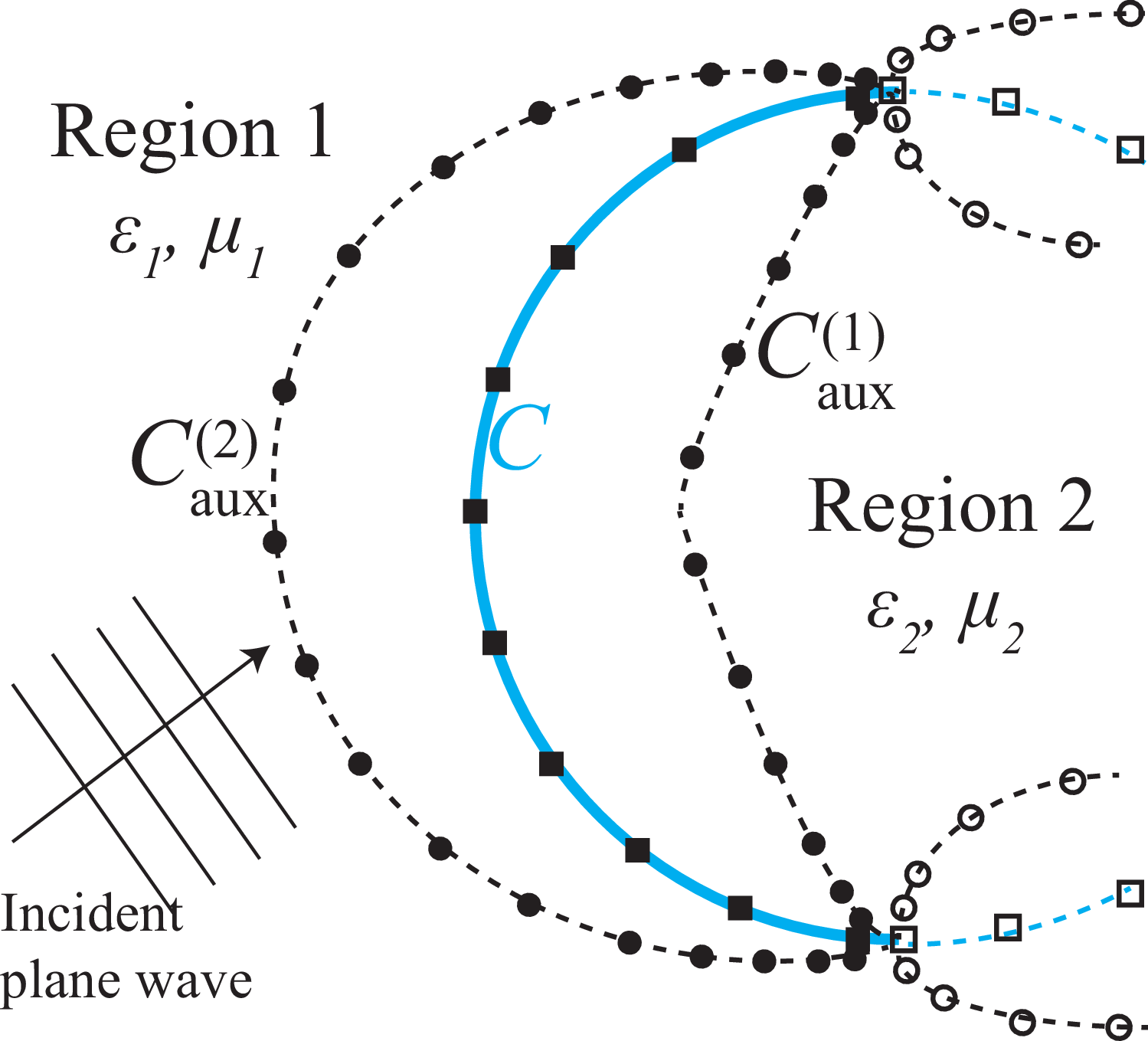}} 
\hfill
\subfigure[]{\includegraphics[width=0.24\textwidth]{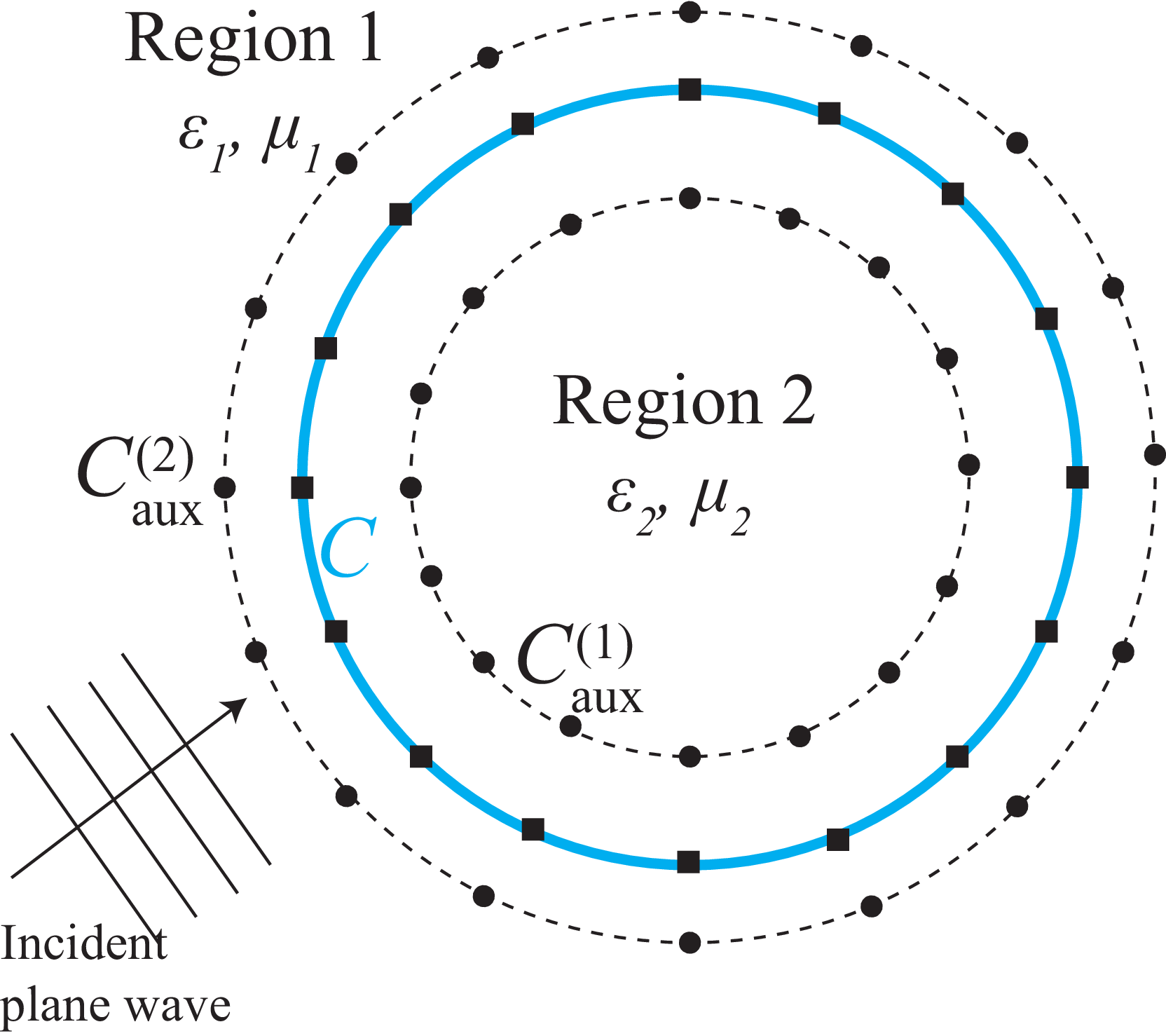}}    
\caption{Application of MAS-SD for the (a) semicircular cylindrical case and (b) the closed cylindrical case. Black circles denote auxiliary sources and black squares denote matching points. In (a), blank circles and squares represent extended auxiliary sources and matching points.}
\label{fig:cylindrical_sch}
\end{figure}

Consider first the semicircular cylindrical metasurface of Fig.~\ref{fig:cylindrical_sch}(a). A special placement of auxiliary sources and matching points is required, including an extension beyond the metasurface edges. 
Specifically, the $N_{m1}$ and $N_{m2}$ auxiliary sources opposite to the metasurface are positioned according to the empirical rules of \cite{leviatan1991analysis}. The $N_e$ auxiliary sources are then extended beyond the edges, as illustrated in Fig.~\ref{fig:cylindrical_sch}(a). Again, in the general case, at each point of the auxiliary surfaces we define a pair of \(z\)-directed electric and magnetic filaments, whereas for metasurfaces with Lorentzian resonators only an electric or a magnetic filament is required at each point, depending on the polarization of the incident wave. The matching points are distributed uniformly along both $C$ (in order to facilitate the evaluation of the second derivatives) and the extended region.

Assuming a Lorentzian resonators metasurface, the resulting linear system is that of Section \ref{sec:finite_planar}, where on the metasurface we use ~\eqref{eq:GSTC_cylindrical} instead of ~\eqref{eq:GSTC_pl_lorentzian}, and the incident field is applied in both $R_1$ or $R_2$.
The described approach is followed for opening angles of the cylindrical section larger than $\pi/3$. If the opening angle is smaller than $\pi/3$, the strategy of \cite{leviatan1991analysis} is adopted, whereby the open boundary is entirely covered with auxiliary sources on both sides rather than extending them only near the edges and the incident field is applied in only one of the regions depending on the source position. This case is not shown here since it 
has been omitted for brevity.

For the closed cylindrical metasurface of Fig.~\ref{fig:cylindrical_sch}(b), the auxiliary surfaces $C^{(1)}_{\rm aux}$ and $C^{(2)}_{\rm aux}$ are chosen as scaled versions of the metasurface boundary, with scaling factors $\sigma^{(1)}_{\rm aux}<1$ and $\sigma^{(2)}_{\rm aux}>1$, respectively with the incident applied to one of the regions depending on the position of the source. The resulting linear system is that of Section \ref{sec:infinite_planar}, with the extended GSTC on the metasurface given by ~\eqref{eq:GSTC_cylindrical} instead of ~\eqref{eq:GSTC_pl_lorentzian}.

\subsection{Polygonal Cylindrical Metasurface}
\label{sec:polygon}
\begin{figure}[htb!]
\centering
\subfigure[]{\includegraphics[width=0.28\textwidth]{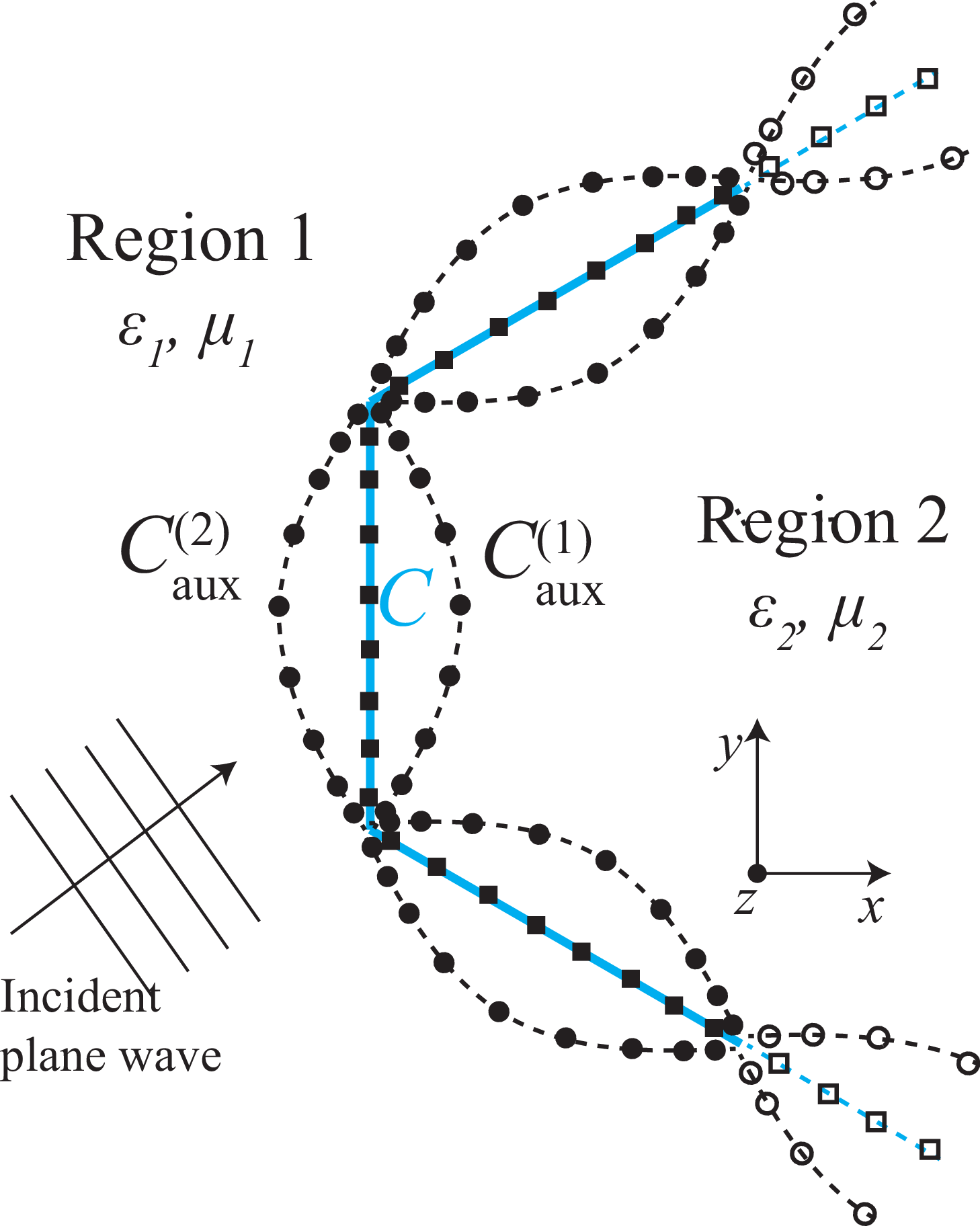}}     
\caption{Application of MAS-SD for an open polygonal cylindrical metasurface. Black circles are auxiliary sources and black squares matching points. Blank circles and squares represent extended auxiliary sources and matching points.}
\label{fig:polygon_sch}
\end{figure}

Now, we consider the open polygonal cylinder shown in Fig. \ref{fig:polygon_sch}(a). Here as well, due to the presence of edges, special treatment is needed for placing the auxiliary sources. Along each straight segment of the polygon, sources are placed with the same approach used for the finite planar metasurface in Section \ref{sec:finite_planar} resulting in $N_{m1}$ and $N_{m2}$ sources placed oppositely on the metasurface. Again, the $N_{e1}$ and $N_{e2}$ sources are expanded in the edges of the polygon resulting in $N_1=N_{m1}+N_{e1}$ on $C^{(1)}_{\rm aux}$ and $N_2=N_{m2}+N_{e2}$ sources on each $C^{(2)}_{\rm aux}$. In the general case, collocated $z$-directed electric and magnetic filaments are placed on the auxiliary surfaces. For metasurfaces with Lorentzian resonators, only one type of filament is used, depending on the incident polarization.

We consider two sets of matching points: \(M_m\) points located on the polygonal metasurface and \(M_e\) (avoiding the singular points) additional points extending beyond its edges. $R_1$ is the region left to the matching points and $R_2$ is the one on the right with the incident field existing in both of them. 
For a metasurface with Lorentzian resonators, the application of the boundary conditions at the matching points leads to a linear system similar to that of Section \ref{sec:finite_planar} with the $\mathbf H_y$ (or $\mathbf E_y$ in case of $\rm TE_z$ polarization) matrices replaced the by $\mathbf H_t$ ($\mathbf E_t$) matrices with the $t$ denoting transverse tangential. 


\subsection{Boundary Condition Error}
To examine the obtained accuracy of MAS-SD, we define the residual error in the boundary conditions as  
\begin{equation}  
e=\frac{||[Z][I]-[V]||_2}{\mathrm {max}\{||[V]||_2, ||[Z][I]||_2\}}  
\end{equation}  
%
For square systems, this error is evaluated at midpoints located midway between each two successive matching points, where the boundary conditions are enforced. For overdetermined systems, the assessment is performed directly at the matching points. The numbers of sources and matching points are increased until an accuracy threshold is met. Hereafter, \mbox{\(e < 5\%\)} is used to determine \mbox{\(N_1, N_2\)} and \mbox{\(M\)}.

\subsection{Second Derivative Calculation}
\label{sec:second_derivative}
At this point, it is important to clarify how the discrete second derivatives of the extended GSTCs (\ref{eq:GSTC_pl_lorentzian}) and (\ref{eq:GSTC_cylindrical}) are computed. Assuming the infinite planar metasurface, let \(F_m\) denote the value of \(E_z\) or \(H_y\) at a given matching point \(m\). Then, for the interior points (i.e., all points except the first \(m=1\) and the last \(m=M\)), we use the central difference approximation
\begin{equation}
\label{eq:intermediate_points}
F''_m \approx \frac{F_{m+1}-2F_m+F_{m-1}}{\Delta y^2}
\end{equation}
%
where  $\Delta y= d_m$. For \(m=1\), the simplest approach is to use the forward second-derivative approximation, namely,
\begin{equation}
\label{fisrt_point}
F''_1 \approx \frac{2F_1-5F_2+4F_3-F_4}{\Delta y^2}.
\end{equation}
For $m=M$, we use the backward approximation
\begin{equation}
\label{eq:last_point}
F''_M \approx \frac{2F_M-5F_{M-1}+4F_{M-2}-F_{M-3}}{\Delta y^2}.
\end{equation}

For the finite planar metasurface, the same equations apply, with the difference that they are enforced only at the \(M_m\) matching points on the metasurface. For an open polygonal cylindrical metasurface, the second derivative is computed separately for each side of the polygon, exactly as applied for the finite planar metasurface, except that for the non-vertical to the $x$-axis sides of the polygon, \(\Delta y\) is replaced by the diagonal distance $\Delta l=\sqrt{\Delta x^2+\Delta y^2}$ between consecutive matching points along each segment. For the circular cylindrical metasurfaces, the same approach as in the polygon case is followed, with \(\Delta y\) replaced by the angular spacing \(\Delta \phi\) between successive matching points.

\section{Numerical Results}
\label{sec:numerical}
This section presents numerical examples validating the proposed MAS-SD formulation for spatially dispersive metasurfaces of increasing geometrical complexity. We begin with infinite planar configurations, including Lorentzian and MIM-based unit cells, and then examine finite planar, polygonal, and smoothly curved metasurfaces. Finally, closed cylindrical structures are considered. For each case, the metasurface response is computed using the extended GSTCs and compared with previously reported results, while additional field profiles are provided when necessary. These examples demonstrate the accuracy, robustness, and flexibility of the proposed MAS-SD for both open and closed spatially dispersive metasurfaces.

\subsection{Infinite Planar Metasurface}
For an infinite planar metasurface with Lorentzian resonators, we simulate the scenario presented in Section IV-B of \cite{gupta_part2}. The metasurface is illuminated by a two-dimensional $\rm TM_z$ Gaussian beam propagating normally toward the surface. The beam is assumed to be invariant along $z$ and confined to the $x$–$y$ plane. The incident electric and magnetic field components are given by ($H_x$ is not needed in the calculations)
\begin{subequations}
\label{eq:gaussian}
\begin{multline}
E_z^{\text{inc}}(x,y) = E_0 
\sqrt{\frac{ j w_0^2}{2x/k_0 +  j w_0^2}}\times\\
\exp\left(-\frac{ j y^2}{2x/k_0 +  j w_0^2}\right)
e^{- j k_0 x}
\end{multline}
\begin{multline}
H^{inc}_y(x,y)=E_0\frac{- j}{\mu_0\omega_0}\sqrt{\frac{ jw_0^2}{\frac{2x}{k_0}+ jw_0^2}}\times\\
\left( \frac{-1}{2x+ jk_0w_0^2}+\frac{2 jk_0y^2}{(2x+ jk_0w_0^2)^2} - jk_0\right)\times\\
\exp\left(-\frac{ j y^2}{2x/k_0 +  j w_0^2}\right)
e^{- j k_0 x}
\end{multline}
\end{subequations}
with $w_0$ the beam waist at $x=0$ and $E_0$ the amplitude of the incident field. For this case, the Lorentzian parameters values in (\ref{eq:GSTC_pl_lorentzian}) (assuming symmetric unit cells), are calculated for operating frequency $f=60~\rm GHz$ and given by
\begin{align*}
\chi_{ee}^{zz}(0) &= -0.0092 +  j\,0.0027, \\
\chi_{mm}^{yy}(0) &= -0.0073 -  j\,0.0062, \\
\xi_{ee,2}^{zz} &= -1.9317\times10^{-5} +  j\,3.8635\times10^{-6}, \\
\xi_{mm,2}^{yy} &= -9.1405\times10^{-5} +  j\,1.8281\times10^{-7}.
\end{align*} 

\begin{figure}[htb!]
\centering
\subfigure[]{\includegraphics[width=0.24\textwidth]{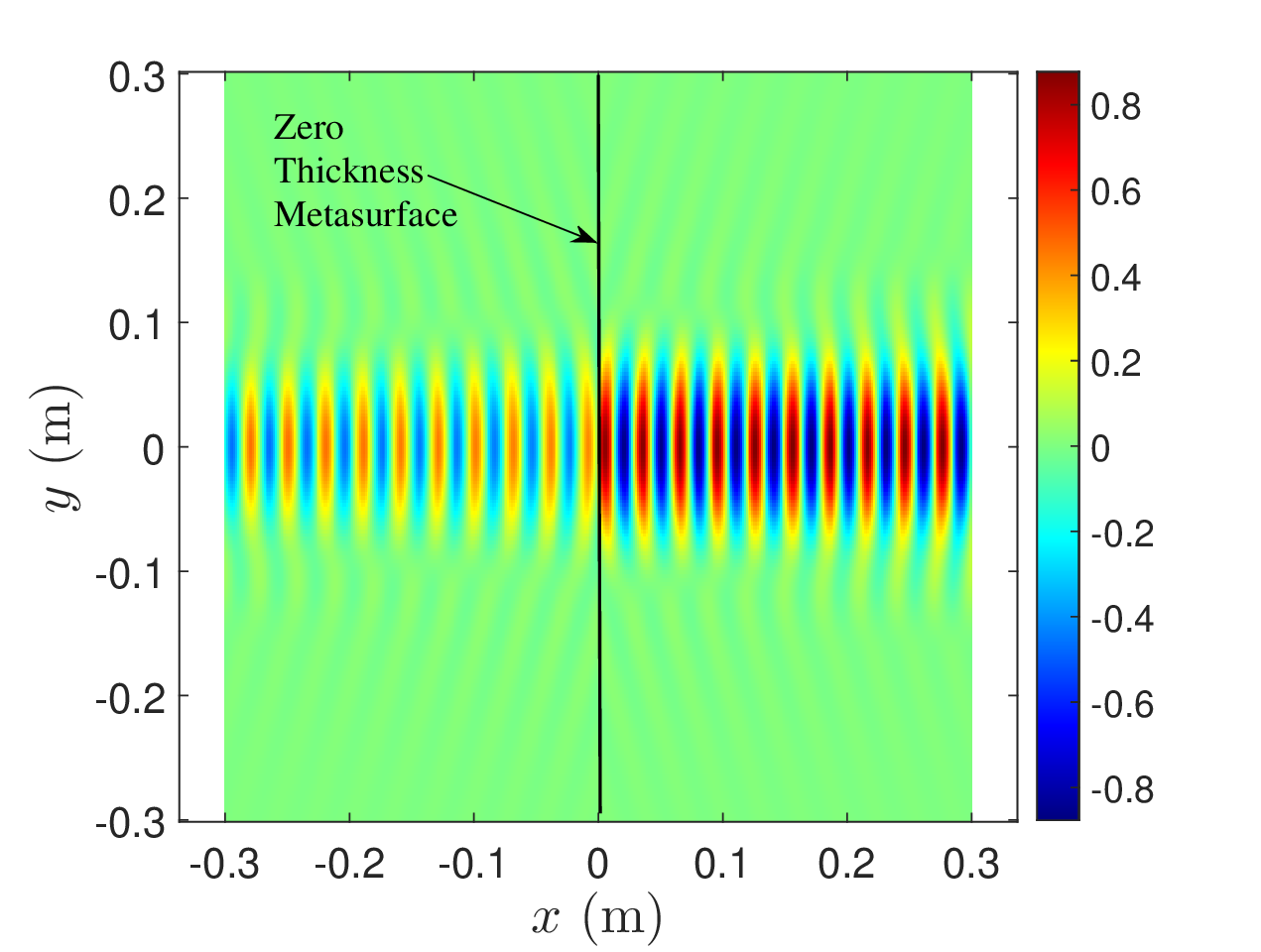}} 
\hfill
    \subfigure[]{\includegraphics[width=0.24\textwidth]{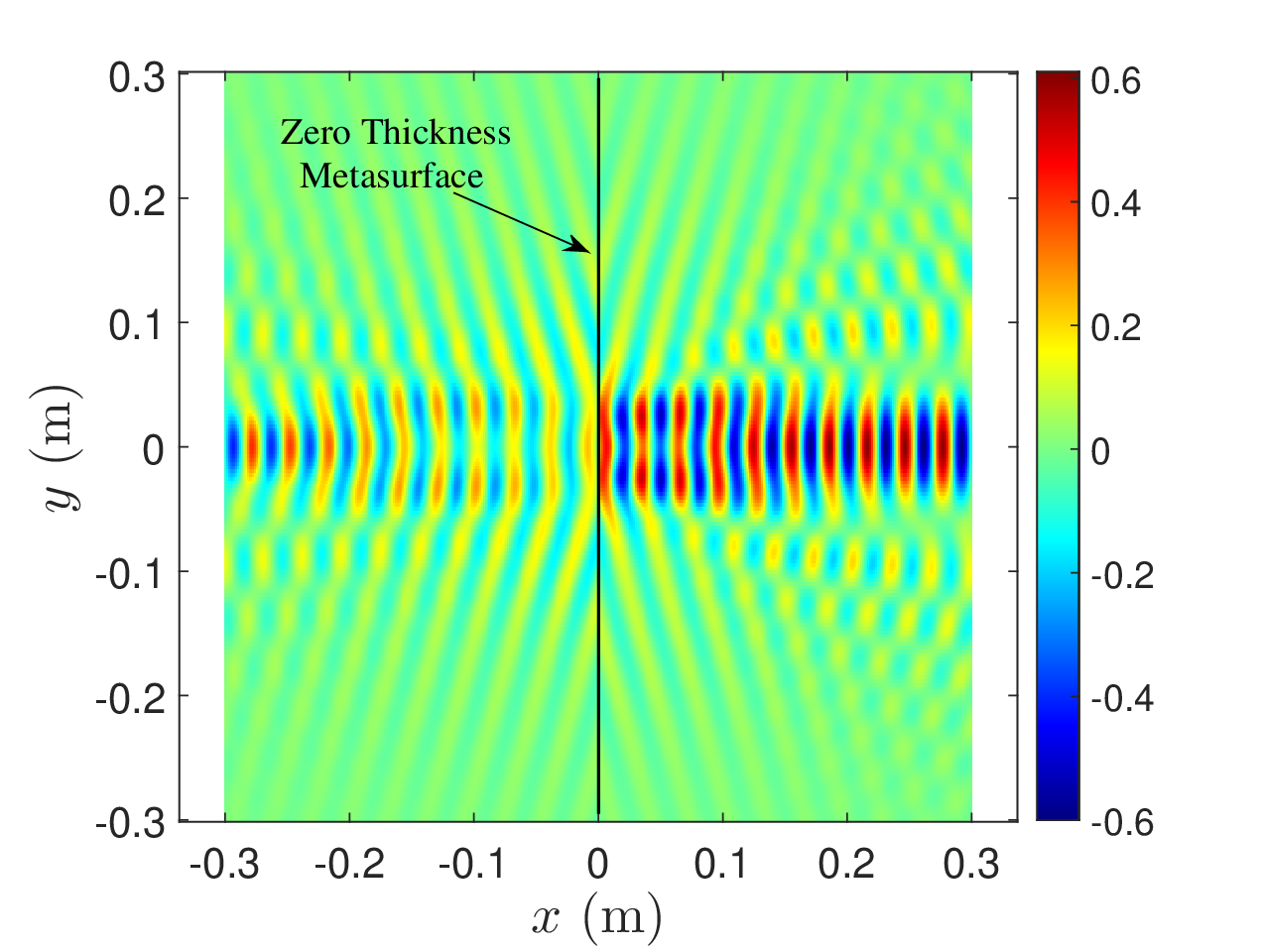}}
    \hfill
    \subfigure[]{\includegraphics[width=0.24\textwidth]{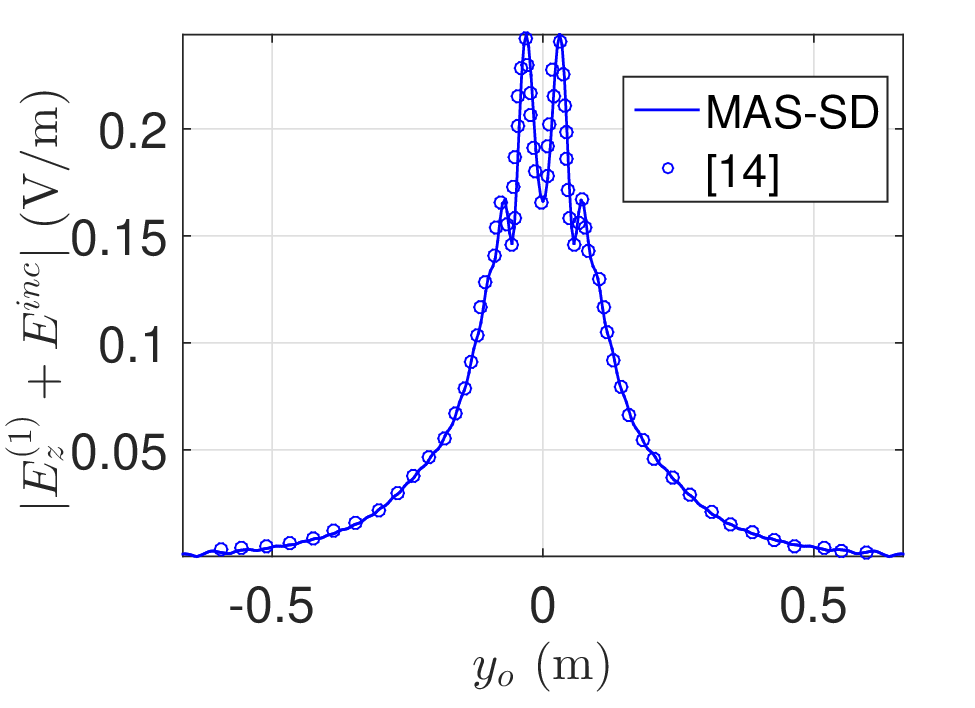}} 
\hfill
    \subfigure[]{\includegraphics[width=0.24\textwidth]{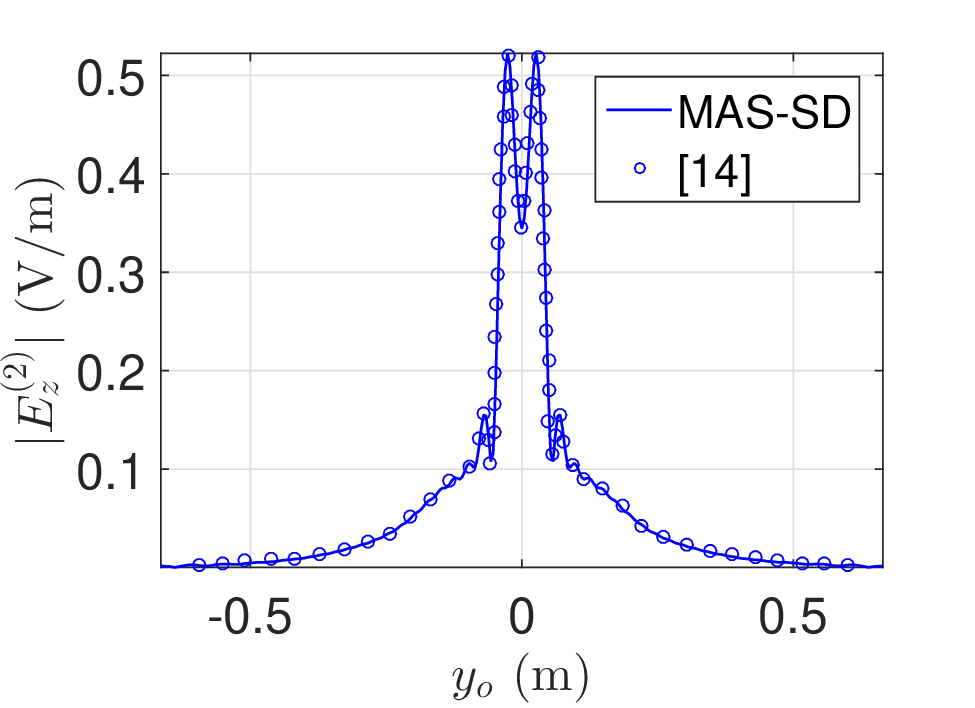}}
\caption{Real part of the scattered electric field, $\mathrm{Re}[E_z]$, for two different widths of the incident Gaussian beam: (a) $w_0=2\lambda$ and (b) $w_0=0.75\lambda$. Magnitude of the total electric field for $w_o=0.75\lambda$: (c) $|E^{(1)}_z+E^{inc}|$ along the line $x=-2\lambda$ and (d) $|E^{(2)}_z|$ along the line $x=2\lambda$.}
\label{fig:infinite_planar}\unskip
\end{figure}

Two scenarios were simulated for two different beamwidths with \(E_0=1\,\mathrm{V/m}\): 
1) a wide beamwidth with \(w_0=2\lambda\), and 
2) a narrow beamwidth with \(w_0=0.75\lambda\). In the MAS-SD implementation, the auxiliary surfaces are placed at $x^{(1)}_{\rm aux}=\lambda/4$ and $x^{(2)}_{\rm aux}=-\lambda/4$. A total of $N_1 = N_2 = 303$ auxiliary sources (electric current filaments) are used on each surface, and the boundary condition is enforced at $M = 303$ matching points on the metasurface, with $d_{\rm aux}=d_m=0.075\lambda$. The resulting linear system is square and, thus, the current vector \(\mathbf I\) is obtained by inverting the impedance matrix \([Z]\). Fig.~\ref{fig:infinite_planar}(a) presents the field distribution obtained using MAS-SD  for $w_0 = 2\lambda$. This field distribution is in very good agreement with the results reported in \cite{gupta_part2}, validating the MAS-SD implementation for this scenario. Fig. \ref{fig:infinite_planar}(b) presents the field distribution obtained by MAS-SD for $w_0 = 0.75\lambda$. The agreement with the reference results is excellent. For further validation of the MAS-SD results against those of \cite{gupta_part2}, we plot in Figs.~\ref{fig:infinite_planar}(c) and (d) the magnitudes of the total electric fields on the lines $x= -2\lambda$ and $x= 2\lambda$ for a narrow beamwidth \(w_0 = 0.75\lambda\) incident Gaussian beam, respectively. The observed agreement with the corresponding results of \cite{gupta_part2} is very good.

\subsection{MIM-Based Infinite Planar Metasurface}

The second infinite planar metasurface case we examine is one whose unit cell consists of a metal–insulator–metal (ΜΙΜ) capacitor printed on a dielectric slab. This case is studied in Section IV-C of \cite{gupta_part2}. The structure is uniform, periodic, and non-magnetic ($\chi^{yy}_{mm}=0$) so it can modeled in terms of a single surface susceptibility, which  for $\rm TM_z$ incidence is given by \cite{gupta_part2}
\begin{equation}
\chi_{ee}^{zz}(k_y) = \chi_{ee,0}^{zz} + \chi_{ee,2}^{zz} k_y^2 .
\end{equation}
This leads to the extended GSTC in the spatial domain
\begin{subequations}
\label{eq:MIM_GSTC}
\begin{equation}
\Delta  H_y= j\omega \varepsilon_0 \chi^{zz}_{ee,0} E_{z,av}- j\omega\varepsilon_0\chi_{ee,2}^{zz} \frac{\partial^2 E_{z,av}}{\partial y^2}
\end{equation}
\begin{equation}
E^{(1)}_z+E^{inc}_z=E^{(2)}_{z}
\end{equation}
\end{subequations}
%
For $f=10~\rm GHz$, we get $\chi^{zz}_{ee,0}=0.0013$ and $\chi^{zz}_{ee,2}=(5.49- j2.98)\times 10^{-7}$. 

\begin{figure}[htb!]
\centering
\subfigure[]{\includegraphics[width=0.24\textwidth]{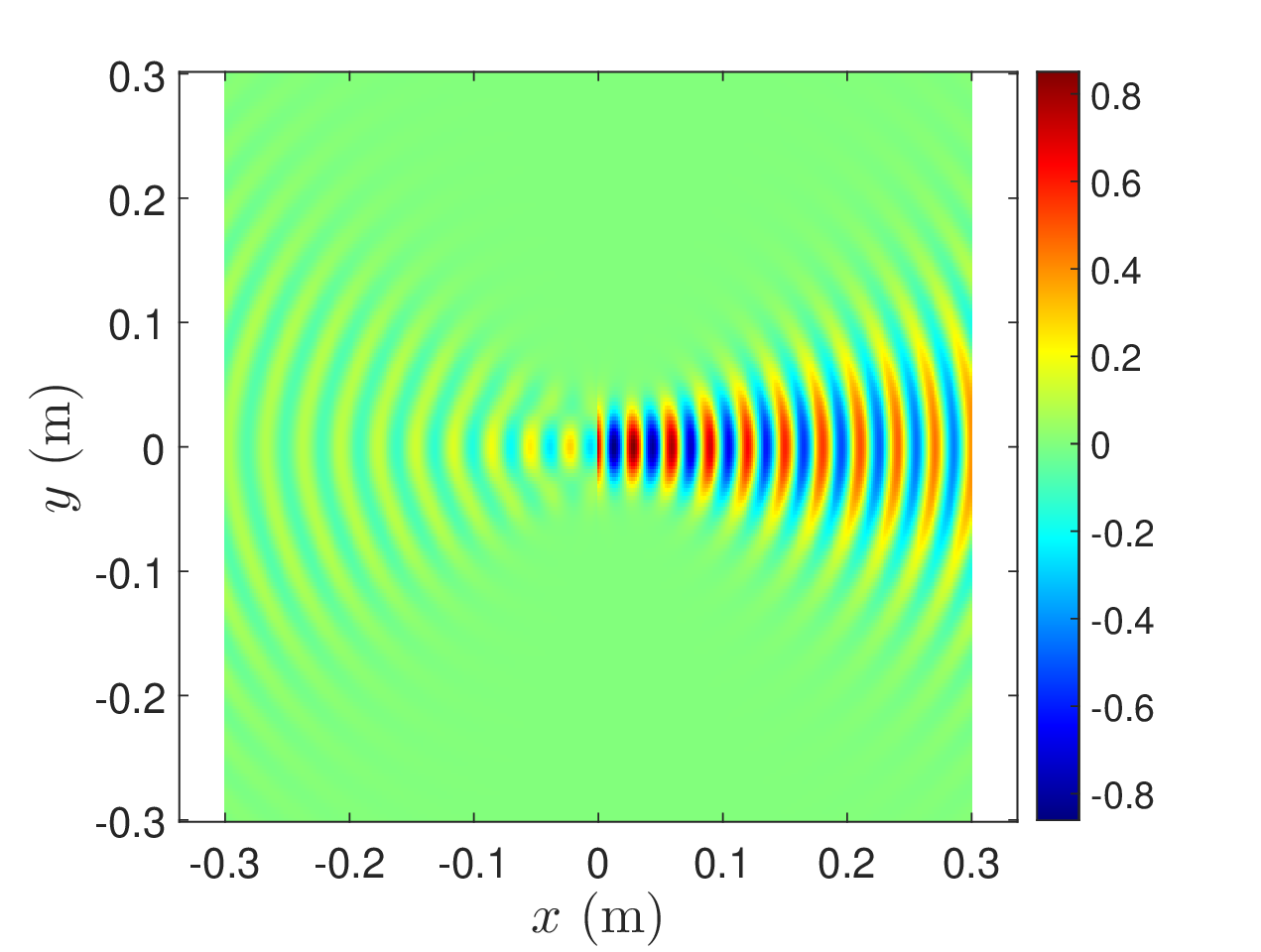}} 
\hfill
    \subfigure[]{\includegraphics[width=0.24\textwidth]{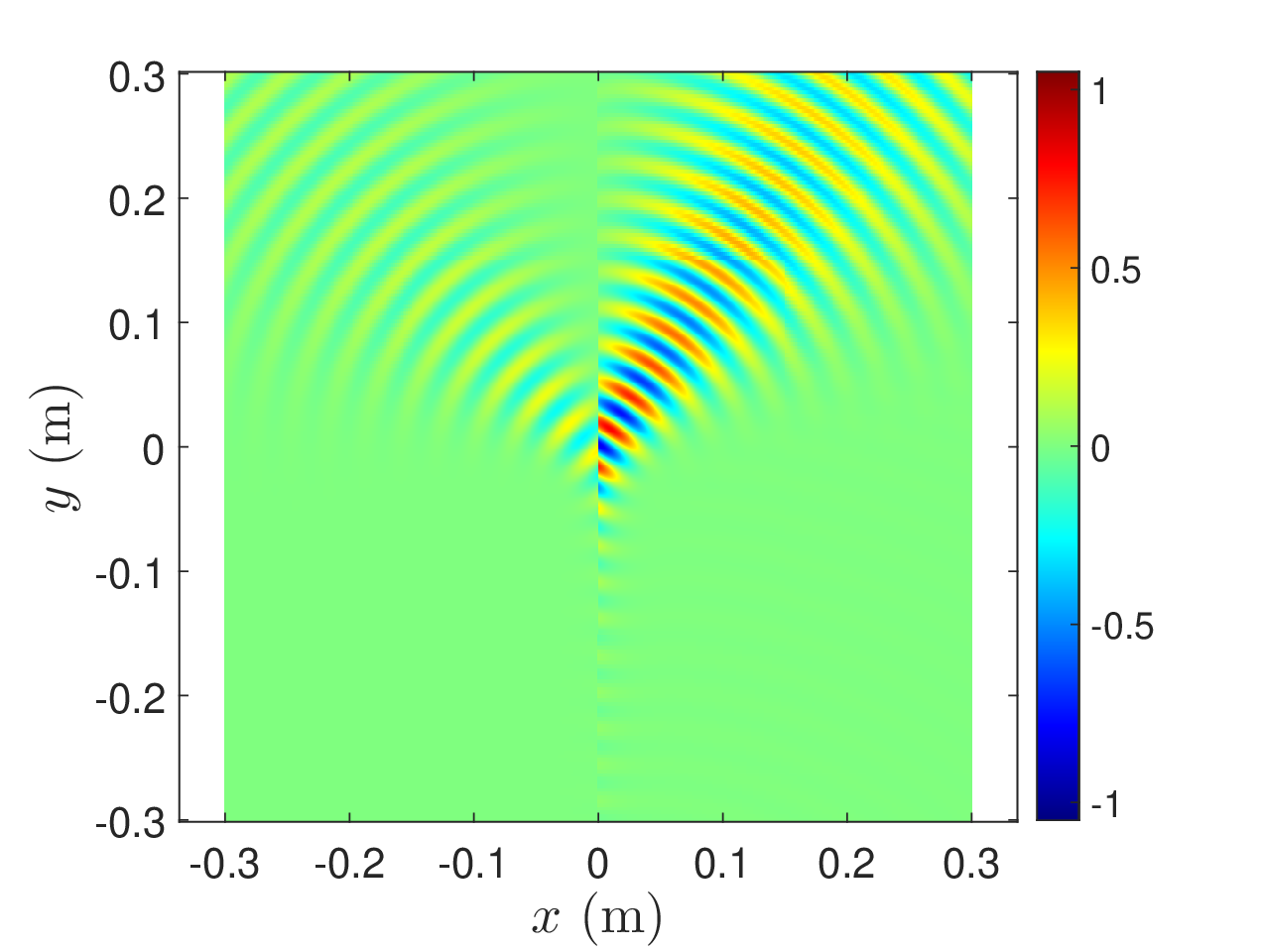}}    
\caption{Real part of the scattered electric field $\mathrm{Re}[E_z]$ for the MIM metasurface with (a) $\theta=0^\circ$ and  (b) $\theta=55^\circ$.}
\label{fig:MIM}
\end{figure}

Using MAS-SD, we simulated this configuration for a Gaussian beam with $w = 0.75\lambda$ (see ~(\ref{eq:gaussian})) under the incidence angles $\phi_{\rm inc} = 0^\circ$ and $\phi_{\rm inc} = 55^\circ$. For both cases $x^{(1)}_{\rm aux}=\lambda/4$ and $x^{(2)}_{\rm aux}=-\lambda/4$, $N_1=N_2=M=151$, with sources and matching points placed on $y$-axis symmetrically with respect to the $x$-axis, with $d_{\rm aux}=d_m=0.075\lambda$. The linear system in matrix form is given by ~(\ref{eq:MIM_IM}). The distribution plots of the total electric field are presented in Fig. \ref{fig:MIM}. In both cases, the obtained fields are very similar to those reported in \cite{gupta_part2}, confirming that the MAS-SD formulation accurately captures the spatially dispersive behavior of the MIM-based metasurface.
\begin{figure*}[htb!]
\begin{align}
\nonumber
\begin{bmatrix}
\mathbf H^{(2)}_y+\frac{ j\omega \varepsilon_0}{2}(\chi^{zz}_{ee,2}\frac{\partial^2 }{\partial y^2}-\chi^{zz}_{ee,0})\mathbf E^{(2)}_z &  \mathbf -H^{(1)}_y+\frac{ j\omega \varepsilon_0}{2}(\chi^{zz}_{ee,2}\frac{\partial^2 }{\partial y^2}-\chi^{zz}_{ee,0})\mathbf E^{(1)}_z \\
\mathbf E^{(2)}_z &  - \mathbf E^{(1)}_z
\end{bmatrix} \begin{bmatrix}
\mathbf I_2 \\ \mathbf I_1
\end{bmatrix}=\\ 
\label{eq:MIM_IM}
\begin{bmatrix}
 \frac{ j\omega \varepsilon_0}{2}(-\chi^{zz}_{ee,2}\frac{\partial^2 }{\partial y^2}+\chi^{zz}_{ee,0})\mathbf E^{inc}-\mathbf H^{inc}\\
\mathbf E^{inc}
\end{bmatrix}
\end{align}
\end{figure*}

\subsection{Finite Planar Metasurface}
\label{sec:finite_planar_num}
The finite planar metasurface we simulated is the same as that presented in Section V-B of \cite{gupta_part2}. It consists of Lorentzian resonators, with operating frequency \(f=60~\mathrm{GHz}\) and length \(l=0.25~\mathrm{m}\). It is placed on the \(yz\)-plane, symmetrically with respect to the \(x\)-axis. The parameter values of the Lorentzian resonators in ~ (\ref{eq:Lorentzian_par}) are: $a_0= 2.25\times 10^9$, $\beta_0=7.593\times 10^9$, $\beta_2=3.578\times 10^{12}$, $\zeta_0=378.7\times 10^9$, $\zeta_2=-18.5\times 10^{15}$, $\chi^{yy}_{mm}=1.84\times 10^{-4}- j5.02\times 10^{-7}$ and $\chi^{zz}_{ee,0}=0$ (not listed parameters are zero). The incident \(\mathrm{TM}_z\) cylindrical wave is due to an electric current filament at \((x_s, y_s)=(-0.03~\mathrm{m}, 0)\) with
\begin{subequations}
\label{eq:TMz_incident}
\begin{equation}
\mathbf{E}^{inc}(x,y)=-\frac{k_0\eta_0}{4}I_{0}H_0^{\left(2\right)}\left(k_0D_{s}\right)\hat{\mathbf{z}},
\end{equation}
\begin{align}
\mathbf{H}^{inc}\left(x,y\right)=&\frac{k_0}{4 j}I_{0}\frac{\hat{\mathbf{x}}\left(y_{s}-y\right)+\hat{\mathbf{y}}\left(x-x_{s}\right)}{D_{s}}H_1^{\left(2\right)}\left(k_0D_{s}\right),
\end{align}
\end{subequations}
and $D_s$ the distance from excitation source to observation point.

In the MAS-SD implementation, as described in Section \ref{sec:finite_planar}, \(N_{m1}=N_{m2}=400\) auxiliary sources were used opposite the metasurface, and \(N_{e1}=N_{e2}=160\) auxiliary sources were extended beyond the edges (half from one edge and the other half from the opposite edge). Additionally, \(M_{m1}=M_{m2}=400\) matching points were used on the metasurface, and \(M_{e1}=M_{e2}=600\) matching points were extended beyond the edges (half from one edge and the other half from the opposite). With these choices, the resulting linear system corresponding to the matrix (\ref{eq:impedance_matrix_finite}) is overdetermined, and we seek the solution with the smallest least-squares error given by
\begin{equation}
\mathbf I=([Z]^*[Z])^{-1}[Z]^*\mathbf V
\label{lease_square_sol}
\end{equation}
where \( ([Z]^*[Z])^{-1}[Z]^* \) is the pseudoinverse of \([Z]\), and \([Z]^*\) denotes the conjugate transpose of \([Z]\).

\begin{figure}[htb!]
\centering
\subfigure[]{\includegraphics[width=0.24\textwidth]{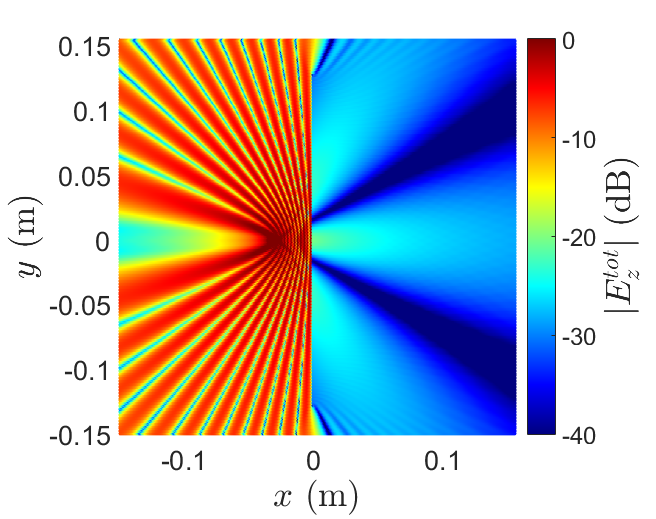}} 
    \hfill
\subfigure[]{\includegraphics[width=0.24\textwidth]{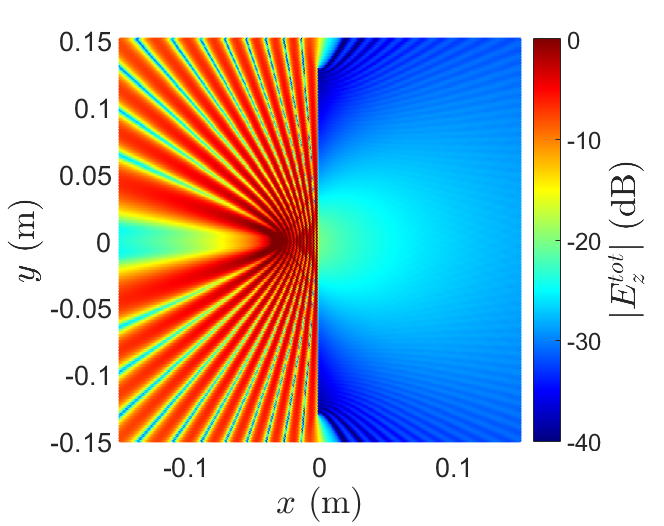}} 
    \hfill
    \subfigure[]{\includegraphics[width=0.24\textwidth]{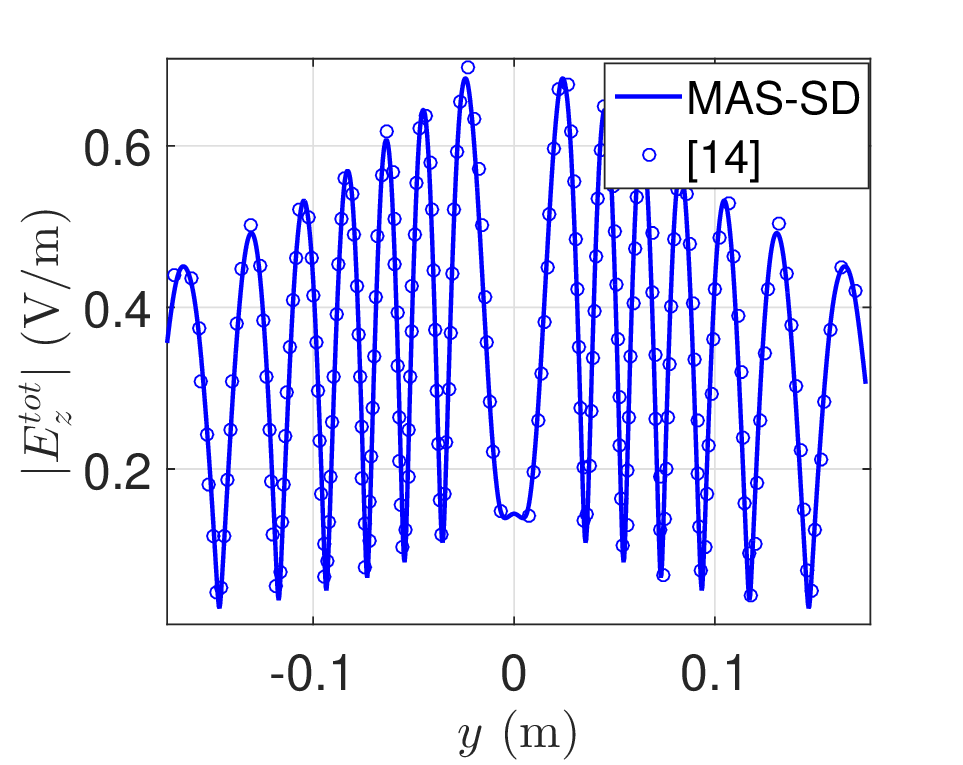}} 
\hfill
    \subfigure[]{\includegraphics[width=0.24\textwidth]{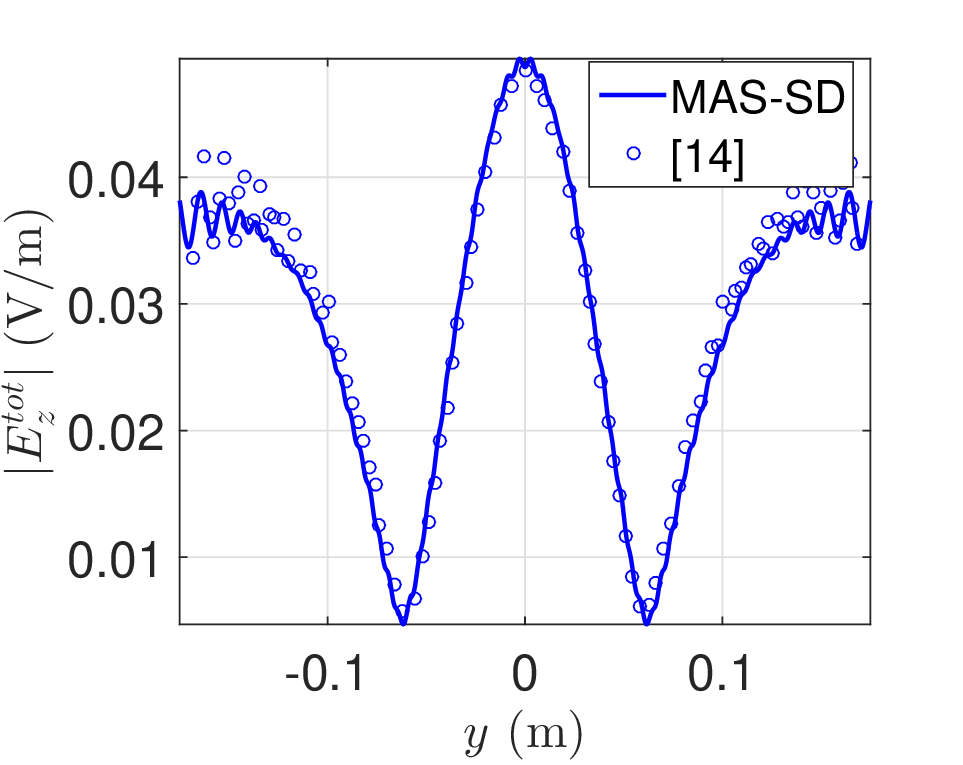}}
\caption{Total electric field magnitude $|E^{\mathrm{tot}}_z|$ for the finite planar metasurface excited by an electric current filament: (a) spatially dispersive metasurface, (b) spatially non-dispersive metasurface. For the spatially dispersive metasurface, total fields along (c) $x=-0.085~\mathrm{m}$ and (d) $x=0.085~\mathrm{m}$ are also shown.}
\label{fig:finite_planar}\unskip
\end{figure}

Fig.~\ref{fig:finite_planar}(a) shows the total electric field magnitude, which can be compared with the corresponding spatially non-dispersive case of Fig.~\ref{fig:finite_planar}(b). 
The magnitude of the total electric field is expressed in dB, normalized to $1~ \rm V/m$.  Comparison with the distributions in \cite{gupta_part2} shows very good agreement confirming the capability of MAS-SD to solve geometries with edges. For additional validation of MAS-SD against the results of \cite{gupta_part2}, Figs.~\ref{fig:finite_planar}(b) and (c) present the magnitude of the total fields on \(x=-0.085~\rm m\) and \(x=0.085~\rm m\). 

\subsection{Semicircular Cylindrical Metasurface}
\label{sec:sem_circ_num}
We now consider the case of a semicircular metasurface with radius \(\rho=0.0342~\mathrm{m}\). This is the second case examined Section V-C in \cite{gupta_part2}. The frequency and the metasurface Lorentzian cells are identical to those used in Section \ref{sec:finite_planar_num}. 
The structure is illuminated by a $\mathrm{TM}_z$-polarized plane wave with $E_0=1~ \rm V/m$ $\phi_{inc}=\pi/6$. Applying the method of Section \ref{sec:cylindrical}, we select \(N_{m1}=N_{m2}=300\), \(N_{e1}=N_{e2}=120\), \(M_m=300\), and \(M_e=240\). These choices lead to an overdetermined linear system, which is solved using the method of Section \ref{sec:finite_planar_num}.

\begin{figure}[htb!]
\centering
\subfigure[]{\includegraphics[width=0.24\textwidth]{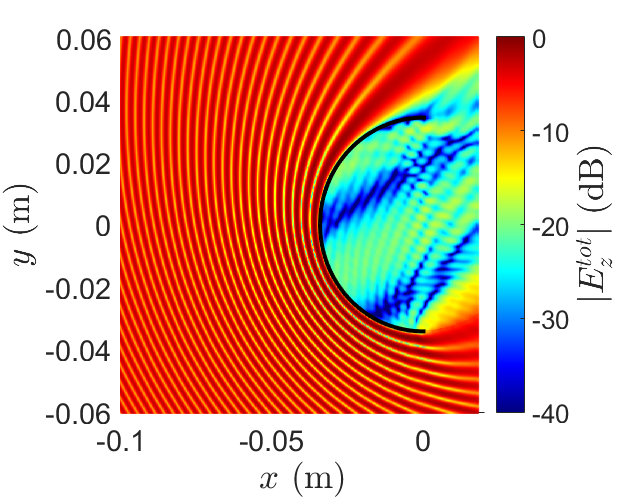}} 
    \hfill
\subfigure[]{\includegraphics[width=0.237\textwidth]{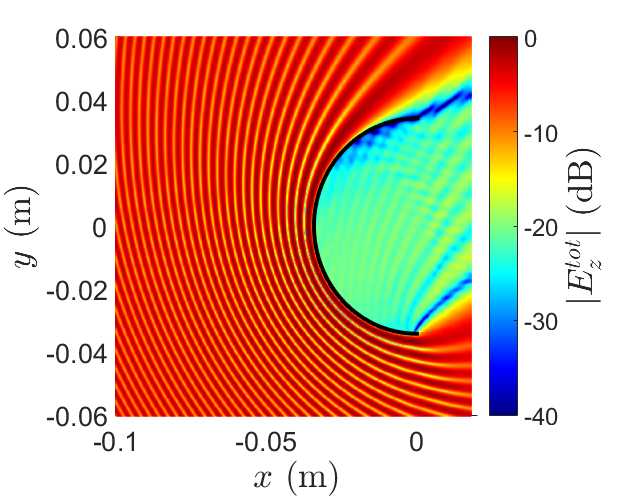}} 
    \hfill
    \subfigure[]{\includegraphics[width=0.24\textwidth]{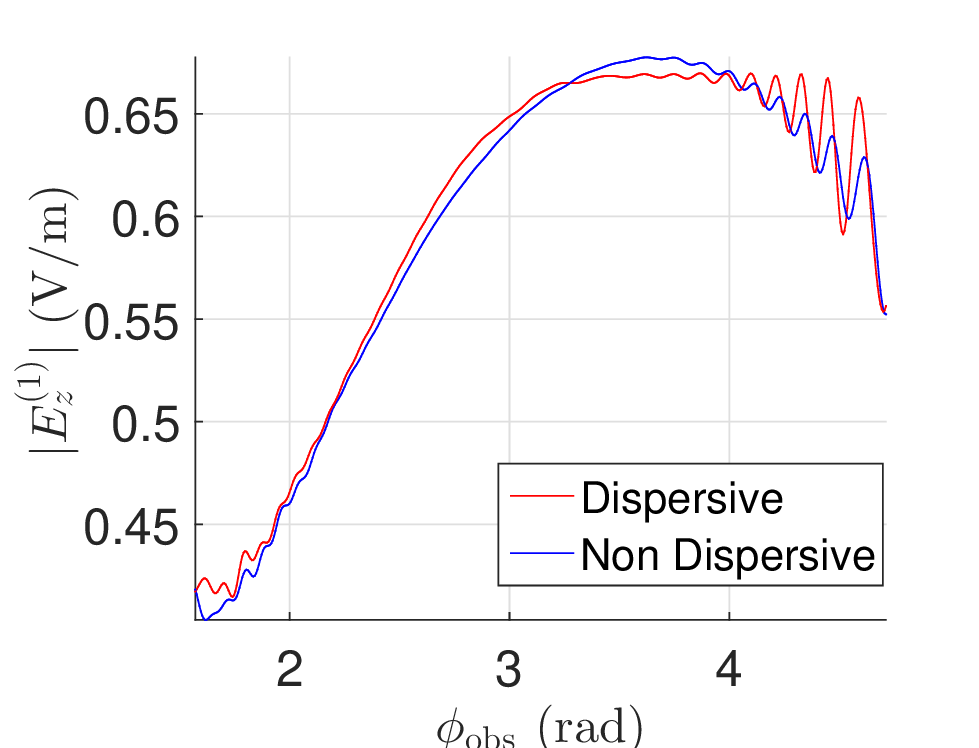}} 
\hfill
    \subfigure[]{\includegraphics[width=0.24\textwidth]{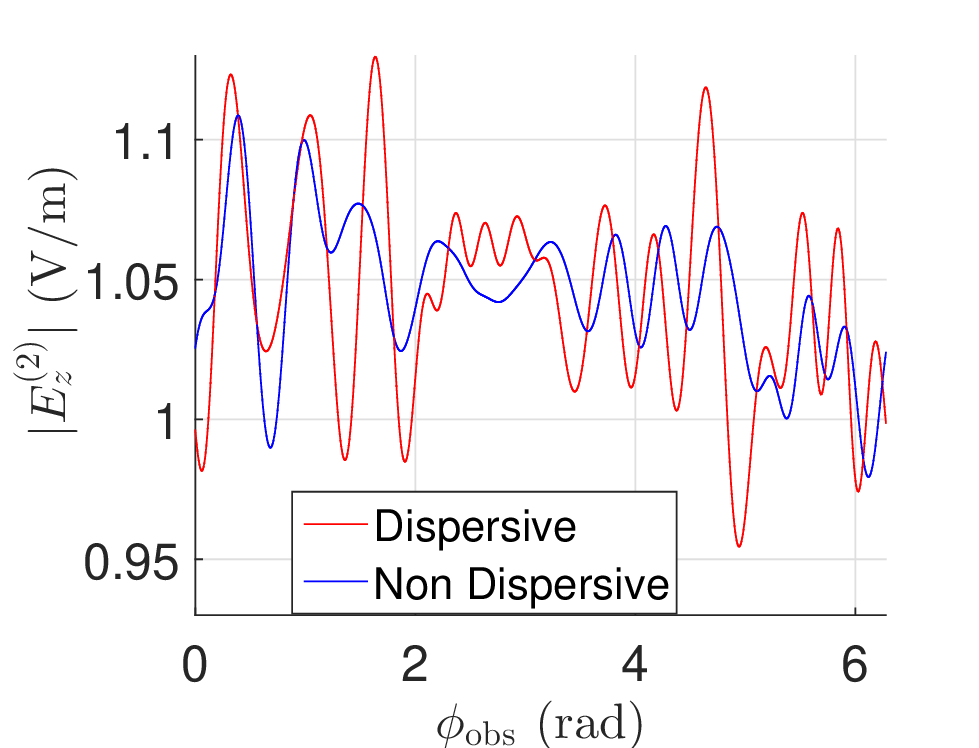}}
\caption{Total electric field magnitude $|E^{\mathrm{tot}}_z|$ for the semicircular cylindrical metasurface excited by a plane wave: (a) spatially dispersive, (b) spatially non-dispersive case. Scattered fields on (c) a semicircle in $R_1$ with $r_o=0.055~\rm m$ centered at $(0,0)$ and (d) a circle in $R_2$ with $r_o=0.01~\rm m$ centered at $(0,0)$.}
\label{fig:semi_circle_num}\unskip
\end{figure}

Fig.~\ref{fig:semi_circle_num}(a) shows the total electric field magnitude, which can be compared with the corresponding spatially non-dispersive case shown in Fig.~\ref{fig:semi_circle_num}(b). The magnitude of the total electric field is expressed in dB, normalized to the maximum field magnitude in $R_1$. Since no additional field profiles are provided in \cite{gupta_part2} for this configuration, we further examine the magnitude of the scattered electric field along two observation contours: 
(i) a semicircle with radius $r_o=0.055~\rm m$ centered at $(0,0)$ located in $R_1$, on the left of the metasurface, and 
(ii) a circle with radius $r_o=0.01~\rm m$ centered at $(0,0)$ positioned in $R_2$ on the right of the metasurface. The corresponding plots are presented in Fig.~\ref{fig:semi_circle_num}(c) and (d) and include both the spatially dispersive and the corresponding spatially non-dispersive case. The comparison highlights the effect of SD on the scattering behavior of the structure, with noticeable differences observed along both observation contours. These results, together with the good agreement of the field distribution plot with \cite{gupta_part2}, support the validity of the MAS-SD for the spatially dispersive semicircular cylindrical metasurface.

\subsection{Closed Circular Cylindrical Metasurface}
\label{sec:circular_metasurface_num}
\begin{figure}[htb!]
\centering
 \subfigure[]{\includegraphics[width=0.48\textwidth]{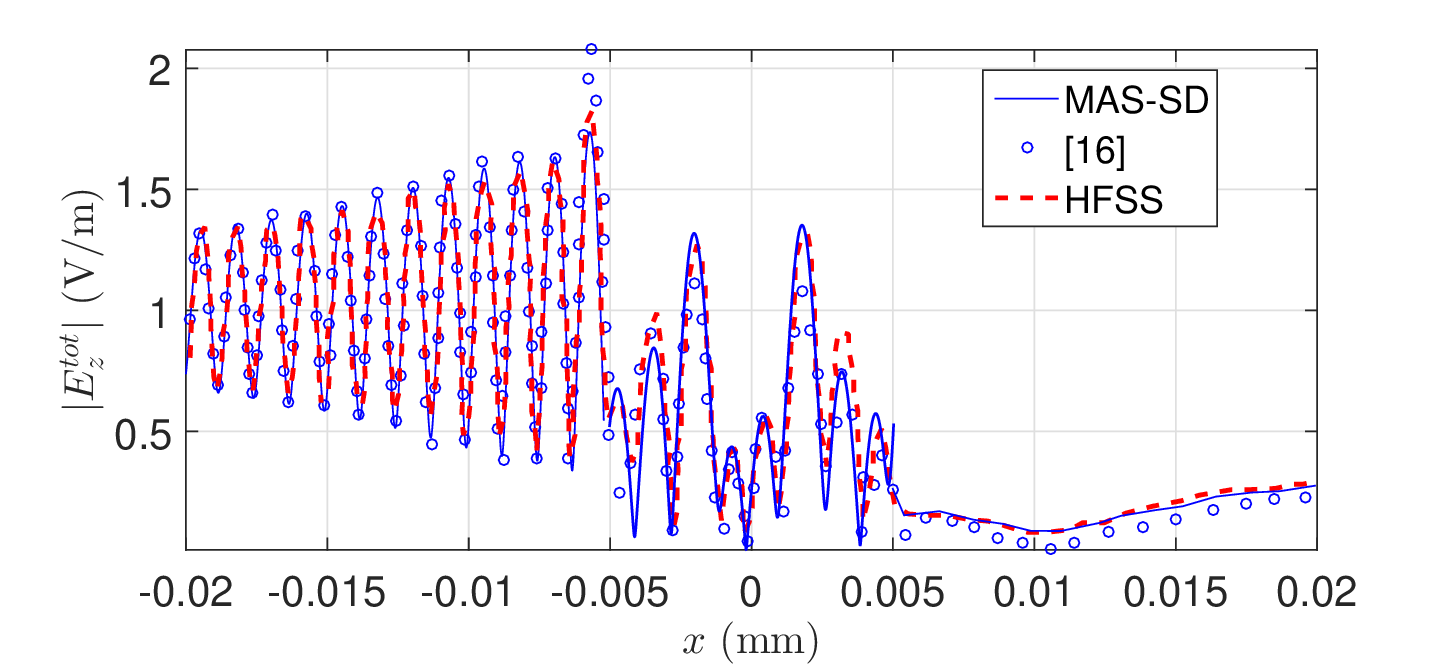}}   
\caption{The magnitude of the total electric field $|E^{\mathrm{tot}}_z|$ along the $x$-axis for the closed circular spatially dispersive metasurface.}
\label{fig:closed_circular}
\end{figure}

Here, we investigate the second case in Section III of  \cite{dugan2023field} of a closed circular cylindrical metasurface of diameter $d=10.25~\rm mm$ excited by a plane wave operating at $f=120~\rm{GHz}$. For the unit cell of this metasurface we use: $a_0=(3.46- j0.191)\times 10^{-3}$, $a_2=0$, $b_2=(-1.02+ j0.0564)\times 10^{-6}$, and $\chi^{yy}_{mm}=1.84\times 10^{-4}- j5.02\times 10^{-7}$ with the rest being zero. Following the method of Section \ref{sec:cylindrical}, we select \(\sigma^{(1)}_{\mathrm{aux}}=0.9\), \(\sigma^{(2)}_{\mathrm{aux}}=1.1\), and \(N_1=N_2=230\) auxiliary sources uniformly placed on the auxiliary surfaces, along with \(M=480\) matching points uniformly distributed on the metasurface. In Fig.~\ref{fig:closed_circular}, the magnitude of the total electric field on the $x$-axis is shown and is found in very good agreement with the reported data in \cite{dugan2023field} (for both IE-GSTC-SD and HFSS).

\subsection{Open Polygonal Cylindrical Metasurface}
\label{sec:polygon_num}
As a last example, we consider an open polygonal metasurface of three sides, each of length \(l=0.05~\mathrm{m}\) (half hexagon). This is the first case of Section V-C in \cite{gupta_part2}. The incident wave is a $\rm TM_z$ plane wave with angle of incidence $\phi_{\rm inc}=60^\circ$.
%
%
The operating frequency is \(f=60~\mathrm{GHz}\), and the elements, and their respective parameter values, are the same to those of the finite planar case in Section \ref{sec:finite_planar_num}.

Applying the method of Section \ref{sec:polygon}, \(N_{m1}=N_{m2}=1200\) auxiliary sources were used, together with \(M_m=1200\) equally spaced matching points. The total number of auxiliary sources extending beyond the edges is \(N_{e1}=N_{e2}=240\) (half from one edge and the other half from the opposite). Additionally, \(M_{e1}=M_{e2}=600\) matching points were extended equidistantly beyond the edges, again with half placed near one edge and the remaining half near the opposite. The overdetermined system is solved by the approach of Section \ref{sec:finite_planar_num}.  

\begin{figure}[htb!]
\centering
\subfigure[]{\includegraphics[width=0.246\textwidth]{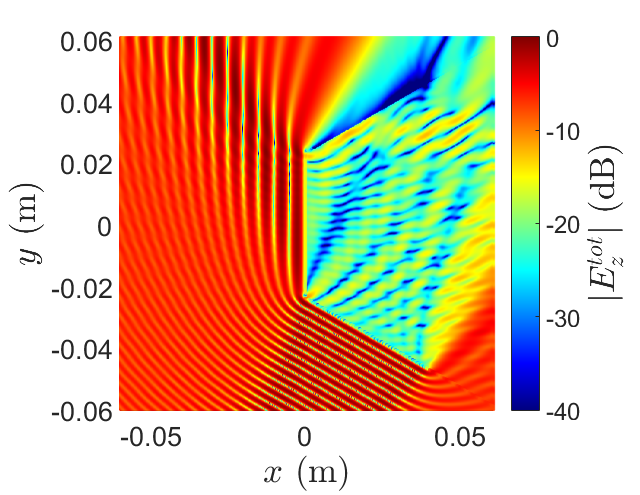}} 
    \hfill
\subfigure[]{\includegraphics[width=0.235\textwidth]{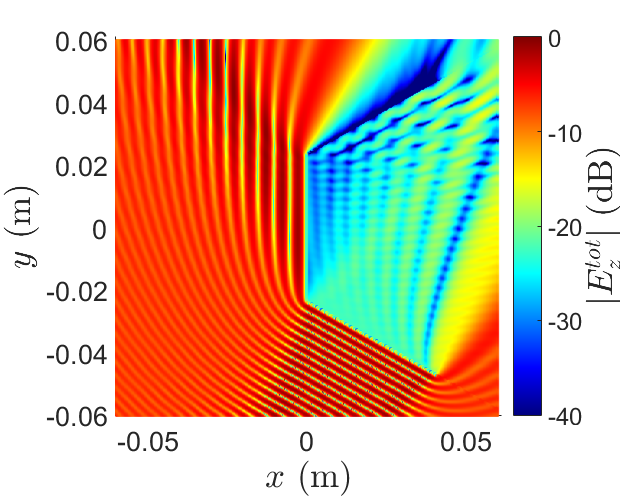}} 
    \hfill
    \subfigure[]{\includegraphics[width=0.24\textwidth]{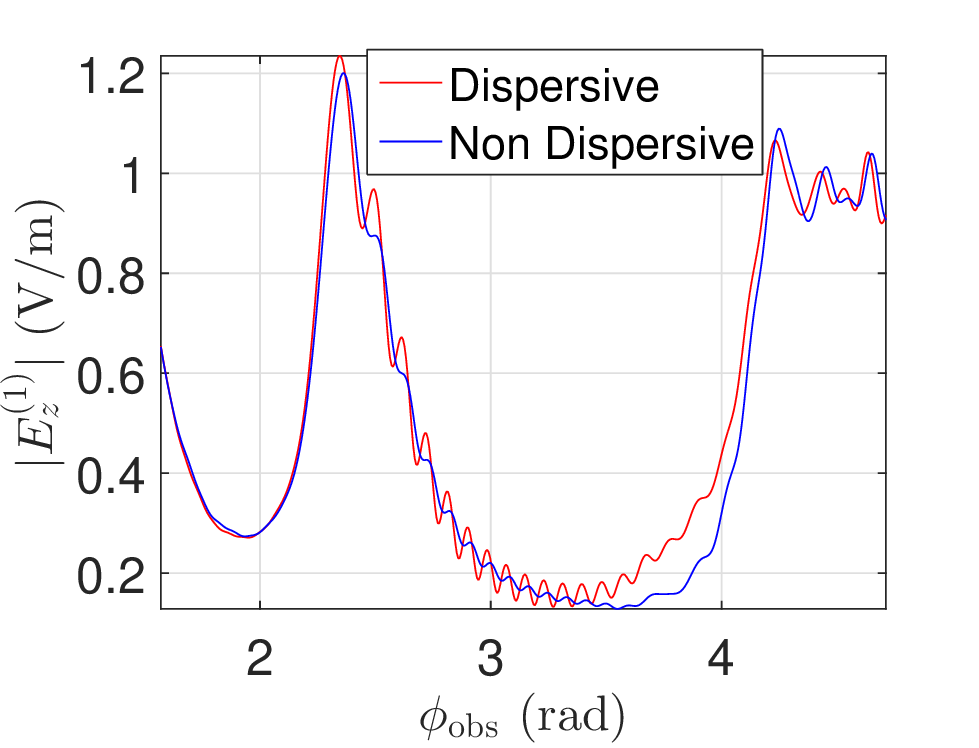}} 
\hfill
    \subfigure[]{\includegraphics[width=0.24\textwidth]{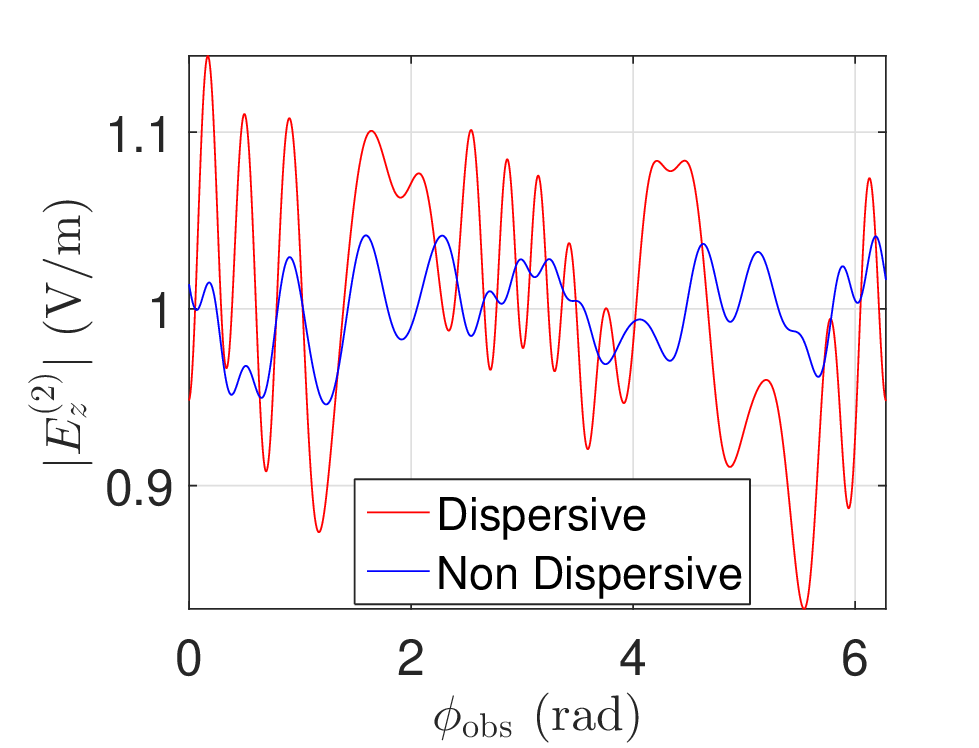}}
\caption{Total electric field magnitude $|E^{\mathrm{tot}}_z|$ for the open polygon cylindrical metasurface excited by a plane wave: (a) spatially dispersive, (b)  spatially nondispersive case. Scattered field profiles on (c) a semicircle in $R_1$ with $r_o=0.055~\rm m$ centered at $(0.02~\rm m,0)$ and (d) a circle in $R_2$ with $r_o=0.01~\rm m$ centered at $(0.02~\rm m,0)$.}
\label{fig:polygon_num}\unskip
\end{figure}

Fig.~\ref{fig:polygon_num}(a) shows the total electric field magnitude, compared with the spatially nondispersive case in Fig.~\ref{fig:polygon_num}(b). The field is expressed in dB and normalized to the maximum value in $R_1$. The results are in very good agreement with \cite{gupta_part2}, confirming the accuracy of MAS-SD. We also examine the field magnitude along two contours: (i) a semicircle of radius $r_o=0.055~\rm m$ in $R_1$ and (ii) a circle of radius $r_o=0.01~\rm m$ in $R_2$. The corresponding plots in Fig.~\ref{fig:polygon_num}(c) and (d) compare the spatially dispersive and nondispersive cases, highlighting the impact of spatial dispersion.

\section{Conclusion}
%

In this work, MAS has been extended to the analysis of spatially dispersive metasurfaces using the extended GSTC formulation. Spatial dispersion is modeled via angle-dependent surface susceptibilities represented by Lorentz-type resonators, leading to differential boundary conditions involving spatial derivatives of the field differences and average fields. These extended GSTCs are incorporated into the MAS, resulting in the MAS-SD formulation. The method is developed for infinite planar, finite planar, half-hexagonal, semicircular, and circular metasurface configurations. Numerical results show very good agreement with previously published data, validating the accuracy of the approach. The proposed method provides a simple, flexible, and computationally efficient meshless alternative to existing techniques. Finally, this work establishes a foundation for extending MAS to space–time modulated metasurfaces, where both frequency and spatial dispersion are present.


\bibliographystyle{IEEEtran}
{\small
\bibliography{mybib_rev_kats}
}

\end{document}